\documentclass[10pt]{article}

\usepackage[T1]{fontenc}
\usepackage[utf8]{inputenc}
\usepackage[english]{babel}
\usepackage{amsmath,latexsym,amsfonts,amsthm,bm,mathtools}
\usepackage{graphicx}
\usepackage[top=1.5in, bottom=1.5in, left=1.25in, right=1.25in]{geometry}
\usepackage{booktabs,url,tikz}
\usepackage[makeroom]{cancel}
\usepackage{dsfont}
\usepackage{relsize}
\usepackage{multirow}

\usepackage{algorithmicx}
\usepackage{algpseudocode}
\usepackage{algorithm}

\newcommand{\Varr}[2]{\mathbb{V}\text{ar}_{\tiny{#1}}\left[ #2\right]}

\newcommand{\Expp}[2]{\mathbb{E}_{\tiny{#1}}\left[ #2\right]}

\newcommand{\Prob}[1]{\mathbb{P}(#1)}

\newcommand{\Probb}[2]{\mathbb{P}_{#2}\left(#1\right)}

\newcommand{\approxx}{\overset{\sim}{=}}

\newcommand{\fuzzyset}[1]{\xi_{\widetilde{#1}}}
\newcommand{\fuzzysetij}[2]{\xi_{\widetilde{#1}_{#2}}}
\newcommand{\indicatorFun}[3]{\mathds{1}_{(#2,#3)}(#1)} 
\newcommand{\digamma}[1]{\Psi\left(#1\right)}

\DeclareMathAlphabet{\mathcall}{OMS}{zplm}{m}{n}

\title{Estimating latent linear correlations \\from fuzzy frequency tables}
\author{Antonio Calcagn\`{i}$^{\ast}$ \\\\
		\footnotesize{\sl University of Padova}\\
		\footnotesize{$\ast$ E-mail: antonio.calcagni@unipd.it}
	}

\date{}

\begin{document}

\maketitle

\begin{abstract}
This research concerns the estimation of latent linear or polychoric correlations from fuzzy frequency tables. Fuzzy counts are of particular interest to many disciplines including social and behavioral sciences, and are especially relevant when observed data are classified using fuzzy categories - as for socio-economic studies, clinical evaluations, content analysis, inter-rater reliability analysis - or when imprecise observations are classified into either precise or imprecise categories - as for the analysis of ratings data or fuzzy coded variables. In these cases, the space of count matrices is no longer defined over naturals and, consequently, the polychoric estimator cannot be used to accurately estimate latent linear correlations. The aim of this contribution is twofold. First, we illustrate a computational procedure based on generalized natural numbers for computing fuzzy frequencies. Second, we reformulate the problem of estimating latent linear correlations from fuzzy counts in the context of Expectation-Maximization based maximum likelihood estimation. A simulation study and two applications are used to investigate the characteristics of the proposed method. Overall, the results show that the fuzzy EM-based polychoric estimator is more efficient to deal with imprecise count data as opposed to standard polychoric estimators that may be used in this context. \\

\noindent {Keywords:} fuzzy frequency; generalized natural numbers; polychoric correlations; fuzzy data analysis
\end{abstract}

\vspace{2cm}

\section{Introduction}\label{sec:1}

The latent linear correlation (LLC) is a measure of bivariate association which is usually adopted when variables are measured at an ordinal level or when data are available in the form of frequency or contingency tables. Because LLCs are quite often used in analysing categorical ordered variables, they are also known as polychoric correlations \cite{olsson1979maximum}. Latent linear or polychoric correlations differ from other measures of association such as Goodman and Kruskal's $\gamma$ or Kendall's $\tau$ in that they are based on a latent continuous parametric model according to which LLCs behave. Given a set of $J$ variables, LLCs are computed pairwise for each pair $j,k$ of variables by considering their joint frequencies $\mathbf N_{R\times C}^{jk} = (n_{11}^{jk},\ldots,n_{rc}^{jk},\ldots,n_{RC}^{jk} )$ over a $R^{jk}\times C^{jk}$ partition space of the variables' domain. The general idea is 
to map the observed counts $\mathbf N_{R\times C}^{jk}$ to the real domain of the bivariate latent density model via the Muthen's thresholds-based approach \cite{muthen1985comparison}, under the constraint that the volumes of the rectangles of the latent density should equal the observed frequencies. In doing so, changing the covariance parameter of the latent model will change the probability distribution over the latent rectangles and hence the probability masses over the cells of $\mathbf N_{R\times C}^{jk}$. Although several parametric models are available for estimating LLCs (e.g., Elliptical, Skew-Gaussian, Copula-based models. See: \cite{lee1988estimation,roscino2006generalization,silvia2012reliability}), the standard formulation based on the Gaussian density with zero means and latent correlations $\mathbf{R}^{jk}_{R\times C}$ is strong enough to be of practical use for many empirical applications (for a recent study, see \cite{jin2017asymptotic,monroe2018contributions}). Some of these include inter-rater agreement \cite{petry2010psychometric}, reliability measurement \cite{zumbo2007ordinal,bonanomi2013polychoric,toth2019applying}, ordinal CFA and SEM \cite{muthen1995technical,yang2010confirmatory,lee2012ordinary}, and polychoric-PCA for dimensionality reduction of discrete data \cite{kolenikov2009socioeconomic}. 

Fuzzy frequency or contingency tables are of particular concern across several disciplines including social, behavioral, and health sciences. Overall, there are two main situations which give rise to fuzzy frequencies, namely when precise data are classified into imprecise categories or, in the opposite case, when fuzzy data are classified into either precise or imprecise categories. Examples of the first case may be found in studies involving socio-economic variables (e.g., income, labor flushes, employment) \cite{yang2017east,da2013migrant}, images or scenes classification \cite{dou2007fuzzy,jadon2001fuzzy}, content analysis \cite{kirilenko2016inter}, reliability analyses \cite{demertzis2018innovative}, evaluation of user-based experiences \cite{lee2005fuzzy}, multivariate analysis of qualitative data \cite{acsan2011biplots,blasius2014visualization}, spatial distributional data \cite{greenacre2013fuzzy}, and human-based risk assessment \cite{dan2017introduction}. By contrast, examples of the second case are most common in studies involving rating scales-based variables such as satisfaction, quality, attitudes, and motivation \cite{calcagni2014dynamic,de2014fuzzy}. What both of these situations have in common is that the $R^{jk}\times C^{jk}$ space constitutes a fuzzy partition and, consequently, observed counts in the classification grid are no longer natural numbers. There have been a number of studies that have tried to deal with fuzzy contingency tables and fuzzy association measures. For instance, \cite{kahraman2004fuzzy} proposed some non-parametric tests generalized to the case of fuzzy data, \cite{grzegorzewski2004distribution} studied fuzzy hypotheses testing based on fuzzy random variables, \cite{denoeux2011maximum} proposed a rank-sum test based on fuzzy partial ordering and introduced a modelization of fuzzy statistical significance test, \cite{hryniewicz2006goodman} generalized the Goodman and Kruskal's $\gamma$ measure to the case of fuzzy observations arranged into contingency tables, \cite{taheri2016contingency} presented the analysis of contingency tables for both fuzzy observations/crisp categories and crisp observations/fuzzy categories cases along with a fuzzy generalization of association measures based on frequencies. Although they differ in some respects, all of them generalize the analysis of contingency tables to the fuzzy case either by the Zadeh's extension principle or by the $\alpha$-cuts based calculus \cite{viertl2011statistical}.

Based on this research stream, this article focuses on estimating latent linear correlations from fuzzy frequency tables, which include both the cases of crisp observations/fuzzy categories and fuzzy observations/fuzzy or crisp categories. Unlike the aforementioned studies, we develop our results by generalizing the standard LLC problem to cope with fuzzy frequencies under the general fuzzy maximum likelihood framework \cite{denoeux2011maximum,quost2016clustering}. In particular, we define the fuzzy frequency table $\mathbf{\widetilde N}_{R\times C}^{jk}$ in terms of fuzzy cardinality and generalized natural numbers first, and then we extend the sample space of the LLC model to deal with fuzzy counts $\tilde n_{11}^{jk},\ldots,\tilde n_{rc}^{jk},\ldots,\tilde n_{RC}^{jk}$. In doing so, the fuzziness of the observations enters the model as a systematic and non-random component while the model's parameters are still crisp (i.e., the estimated latent correlation matrix $\mathbf{\widehat R}^{jk}_{R\times C}$ is a non-fuzzy quantity). This offers an attractive solution to the problem of estimating LLCs with fuzzy information, with the additional benefit that statistical models that uses the LLCs statistic as input data (e.g., CFA, PCA, SEM) do not need any further generalization to cope with fuzzy data.

The reminder of this article is structured as follows. Section \ref{sec:2} introduces the concept of fuzzy frequency through fuzzy cardinalities and generalized natural numbers. Section \ref{sec:3} describes the fuzzy LLCs model and its characteristics in terms of parameters estimation and interpretation. Section \ref{sec:4} reports the results of a simulation study performed to assess the finite sample properties of the fuzzy LLCs model as compared with standard defuzzification-based estimation methods. Section \ref{sec:5} describes the application of the proposed method to two real case studies and section \ref{sec:6} concludes the article by providing final remarks and suggestions for further extensions of the current findings. All the materials like algorithms and datasets used throughout the article are available to download at \url{https://github.com/antcalcagni/fuzzypolychoric/}.

\section{Fuzzy frequencies}\label{sec:2}

\subsection{Preliminaries}\label{sec:2_1}
A \textit{fuzzy subset} $\tilde{A}$ of a universal set $\mathcall A \subset \mathbb R$ can be defined by means of its characteristic function $\fuzzyset{A}:\mathcall{A}\to [0,1]$. It can also be expressed as a collection of crisp subsets called $\alpha$-sets, i.e. $\tilde{A}_\alpha = \{x \in \mathcall A: ~\fuzzyset{A}(x) > \alpha \}$ with $\alpha \in (0,1]$. If the $\alpha$-sets of $\tilde{A}$ are all convex sets then $\tilde{A}$ is a {convex fuzzy set}. The \textit{support} of $\tilde{A}$ is ${A}_{0} = \{x \in \mathcall A: ~\fuzzyset{A}(x) > 0 \}$ and the \textit{core} is the set of all its maximal points ${A}_{1} = \{x \in \mathcall A: ~\fuzzyset{A}(x) = \max_{z \in \mathcall A}~ \fuzzyset{A}(z) \}$. In the case $\max_{x\in \mathcall A} \fuzzyset{A}(x) = 1$ then $\tilde{A}$ is a {normal} fuzzy set. If $\tilde{A}$ is a normal and convex subset of $\mathbb R$ then $\tilde{A}$ is a \textit{fuzzy number} (also called fuzzy interval). The quantity $l(\tilde A) = \sup {A}_0 - \inf {A}_0$ is the \textit{length} of the support of the fuzzy set $\tilde A$. The \textit{simple cardinality} of a fuzzy set $\tilde A$ is defined as $|\tilde A| = \int_{\mathcall A}\fuzzyset{A}(x)~dx$. Given two fuzzy sets $\tilde A, \tilde{B}$, the \textit{degree of inclusion} of $\tilde{A}$ in $\tilde{B}$ is $\epsilon_{\tilde{A}\tilde{B}} = \left|\min_{x\in\mathcall A}\left(\fuzzyset{A}(x),\fuzzyset{B}(x)\right)\right| \big/ \max(1,|\tilde A|)$, with $\epsilon_{\tilde{A}\tilde{B}}\in[0,1]$. The case $\epsilon_{\tilde{A}\tilde{B}}=1$ indicates that $\tilde A$ is fully included in $\tilde B$. The class of all normal fuzzy numbers is denoted by $\mathcall F(\mathbb{R})$. Fuzzy numbers can conveniently be represented using parametric models that are indexed by some scalars. These include a number of shapes like triangular, trapezoidal, gaussian, and exponential fuzzy sets \cite{hanss2005applied}. A relevant class of parametric fuzzy numbers are the so-called LR-fuzzy numbers \cite{dubois2012fundamentals} and their generalizations \cite{calcagni2014non,toth2019applying}. The \textit{trapezoidal fuzzy number} is one of the most common fuzzy set used in many applications and it is parameterized using four parameters as follows:
\begin{equation}\label{eq1}
	\fuzzyset{A}(x) = \indicatorFun{x}{c_1}{c_2} + \bigg(\frac{x-x_l}{c_1-x_l}\bigg) \indicatorFun{x}{x_l}{c_1} + \bigg(\frac{x_u-x}{x_u-c_2}\bigg)\indicatorFun{x}{c_2}{x_u}
\end{equation}
with $x_l,x_u,c_1,c_2 \in \mathbb R$ being lower, upper bounds, and first and second modes, respectively. The symbol $\indicatorFun{x}{a}{b}$ denotes the indicator function in the interval $(a,b)$. Interestingly, the trapezoidal fuzzy set includes the \textit{triangular} (if $c_1=c_2$) and \textit{rectangular} (if $x_l=c_1, c_2=x_u$) fuzzy sets as special cases. A \textit{degenerated} fuzzy number $\mathring{A}$ is a particular fuzzy set with $\fuzzyset{A}(c)=1$ and $\fuzzyset{A}(x)=0$ for $x\neq c$, $x\in \mathcall A$. Note that rectangular and degenerated fuzzy numbers can be adopted to represent crisp categories and crisp observations, respectively. When a probability space is defined over $\mathcall A$, the \textit{probability} of a fuzzy set $\tilde A$ can be defined as $\Prob{\tilde{A}} = \int_{\mathcall A} \fuzzyset{A}(x)d\mathbb{P}$ (Zadeh's probability). Similarly, the joint probability of two fuzzy sets is $\Prob{\tilde{A}\tilde{B}} = \int_{\mathcall A} \fuzzyset{A}(x)\fuzzyset{B}(x)d\mathbb{P}$ under the rule $\xi_{\tilde{A}\tilde{B}}(x)= \fuzzyset{A}(x)\fuzzyset{B}(x)$ (independence of fuzzy sets) \cite{zah1968probability}. 

\subsection{Fuzzy granules}\label{sec:2_2}

Let $\mathcall S = \{\tilde{A}_1,\ldots,\tilde A_i,\ldots,\tilde{A}_I\}$ be a sample of $I$ fuzzy or non-precise observations with $\tilde{A}_i$ being a fuzzy number as defined by Eq. \eqref{eq1}. Then the interval $\mathcall R(\mathcall S) = [r_{0},r_{1}]\subset \mathbb R$ is the \textit{range} of the fuzzy sample where $r_0 = \min \{A^\dagger_{0_1},\ldots,A^\dagger_{0_I}\}$ and $r_1 = \max \{A^\dagger_{0_1},\ldots,A^\dagger_{0_I}\}$, with $A^\dagger_{0_i}$ being the infimum of the support set $A_{0_i}$ computed for the $i$-th fuzzy observation. A collection $\mathcall G = \{\tilde{G}_1,\ldots,\tilde{G}_c,\ldots,\tilde G_C\}$ of $C$ fuzzy sets is a \textit{fuzzy partition} of $\mathcall R(\mathcall S)$ if the following two properties hold (i) $\max_{i=1,\ldots,I} l(\tilde A_i) \leq \min_{c=1,\ldots,C} l(\tilde C_c)$ and (ii) $\sum_{c=1}^{C} \fuzzyset{G_c}(x) = 1$ (Ruspini's partition) \cite{bodjanova2008,gil1984comparison}. The fuzzy sets in $\mathcall G$ are also called \textit{granules} of $\mathcall R(\mathcall S)$. The evaluation of the amount of fuzzy observations in a granule $\tilde{G}_c$ is called \textit{cardinality} (scalar or fuzzy) and can be used to compute fuzzy frequencies or counts for a partition $\mathcall G$ given a sample $\mathcall S$. Figure \ref{fig1} (leftside panels) shows an example of fuzzy granulation for both fuzzy and crisp observations.

\subsection{Fuzzy counts as generalized natural numbers}\label{sec:2_3}
Let $\mathbf{ \tilde x}_j = \{\tilde{x}^j_{1},\ldots,\tilde{x}^j_{i},\ldots,\tilde{x}^j_{I}\}$ and $\mathbf{ \tilde x}_k = \{\tilde{x}^k_{1},\ldots,\tilde{x}^k_{i},\ldots,\tilde{x}^k_{I}\}$ be two samples of fuzzy observations and $\mathbf{\tilde g}_j = \{\tilde g^j_{1},\ldots,\tilde g^j_{c},\ldots,\tilde{g}^j_{C}\}$ and $\mathbf{\tilde g}_k = \{\tilde g^k_{1},\ldots,\tilde g^k_{r},\ldots,\tilde{g}^k_{R}\}$ be two fuzzy partitions of the domains $\mathcall R(\mathbf{ \tilde x}_j)$ and $\mathcall R(\mathbf{ \tilde x}_k)$. Given a pair of granule $(\tilde g_r, \tilde g_c)$, a fuzzy or imprecise count for the joint sample $(\mathbf{ \tilde x}_j,\mathbf{ \tilde x}_k)$ is a fuzzy set $\tilde n_{rc}^{jk}$ with membership function $\fuzzysetij{n}{rc}^{jk}: \mathbb N_{0} \to [0,1]$. As it is defined over natural numbers, a fuzzy count is a finite \textit{generalized natural number} for which extended operations are available (e.g., addition, multiplication) \cite{wygralak1999questions}. Analogously to fuzzy intervals, the class of all fuzzy counts is denoted as $\mathcall F(\mathbb N_{0})$. There are different choices for the computation of $\fuzzysetij{n}{rc}^{jk}$ (e.g., see: \cite{hryniewicz2006goodman,taheri2016contingency,viertl2011statistical,delgado1993inductive,trutschnig2008strong}). In this contribution, we will follow the findings of \cite{bodjanova2000generalized} and \cite{bodjanova2008} which are based on Zadeh's fuzzy counting functions \cite{zadeh1983computational} and fuzzy cardinalities \cite{casasnovas2003axiomatic}. More precisely, let $$\boldsymbol{\epsilon}_{rc}^{jk} = \left(\epsilon_{rc_1}^{jk},\ldots,\epsilon_{rc_i}^{jk},\ldots,\epsilon_{rc_I}^{jk}\right)$$
be the vector of joint degrees of inclusion for the $rc$-th granule where
\begin{align*}
	&\epsilon_{rc_i}^{jk} = \min\left( \epsilon_{r_i}^j,\epsilon_{c_i}^k \right)\\
	&\epsilon_{r_i}^{j} = \big|\min_{x\in \mathcall R(\mathbf{ \tilde x}_j)} (\fuzzysetij{x}{ji}(x),\fuzzysetij{g}{r}(x))~\big|~ \big/ \max(1,|\tilde x_{i}^j|) \\
	&\epsilon_{c_i}^{k} = \big|\min_{x\in \mathcall R(\mathbf{ \tilde x}_k)} (\fuzzysetij{x}{ki}(x),\fuzzysetij{g}{c}(x))~\big|~ \big/ \max(1,|\tilde x_{i}^k|)
\end{align*}
with $|.|$ being the simple cardinality according to the definition given in Sect. \ref{sec:2_1}. For $n\in \mathbb N_0$, the fuzzy count is as follows:
\begin{equation}\label{eq2}
	\fuzzysetij{n}{rc}^{jk}(n) = \min\left( \mu_\text{\tiny FLC}(n), \mu_\text{\tiny FGC}(n) \right)
\end{equation}
with $\mu_\text{\tiny FLC}(n)$ and $\mu_\text{\tiny FGC}(n)$ being the output of the Zadeh's fuzzy counting functions \cite{zadeh1983computational}. The following calculus can be used for $\mu_\text{\tiny FLC}(n)$ and $\mu_\text{\tiny FGC}(n)$. First, compute the square matrix of differences $\mathbf Z_{I\times I} = \left(\boldsymbol{\epsilon}_{rc}^{jk}\mathbf 1_I^T - \mathbf 1_I(\boldsymbol{\epsilon}_{rc}^{jk})^T\right)$, with $\mathbf 1_I$ being a $I\times 1$ vector of all ones. Then, for each $i=1,\ldots,I$ the vector $\mathbf z_{I\times 1}$ is computed, with $z_i = \mathbf 1_I^T\mathcall H(\mathbf Z_{,i})$ and $\mathcall H(x)$ being the Heaviside step-function defined by $\mathcall H(x) := \{0 ~\text{if}~ x<0~,~ 1 ~\text{if}~ x\geq0\}$. The vector $\mathbf z = (z_1,\ldots,z_i,\ldots,z_I)$ contains the sums of the output of the Heaviside function applied column-wise on $\mathbf Z$. Finally, for $n=0,1,2,\ldots,I$ the Zadeh's counting functions are as follows:
\begin{align}\label{eq3}
	& \mu_\text{\tiny FGC}(n) = \max\left( \mathcall H(\mathbf z-n)\odot \boldsymbol{\epsilon}_{rc}^{jk} \right)\\\nonumber
	& \mu_\text{\tiny FLC}(n) = 1-\max\left( \mathcall H(\mathbf z-n+1)\odot \boldsymbol{\epsilon}_{rc}^{jk} \right)\nonumber
\end{align}
where $\odot$ is the element-wise product whereas $\mathcall H(x)$ is the Heaviside function defined as above. Thus, the membership function of $\tilde n_{rc}^{jk}$ is defined as the minimum between the degree of possibility that at least $n$ elements from $(\mathbf{ \tilde x}_j,\mathbf{ \tilde x}_k)$ are included in the $rc$-th granule (FGC count) and the degree of possibility that at most $n$ elements are included in the $rc$-th granule (FLC count). By applying Eqs. \eqref{eq2}-\eqref{eq3} for each pair of granules $(\tilde g_1,\tilde{g}_1),\ldots,(\tilde g_r,\tilde{g}_c),\ldots,(\tilde g_R,\tilde{g}_C)$ one obtains the fuzzy frequency matrix $\mathbf{\widetilde N}_{R\times C}^{jk}$. Note that the resulting fuzzy set $\fuzzysetij{n}{rc}$ may not be normal, i.e. $\max_{n} \fuzzysetij{n}{rc}(n) \leq 1$, and a post-hoc normalization should be applied if normal fuzzy sets were needed. Table \ref{alg1} summarizes the algorithm for computing fuzzy frequencies for a given pair of fuzzy variables.

\begin{figure}[!h]
	\hspace{-1cm}
	\resizebox{17cm}{!}{\input{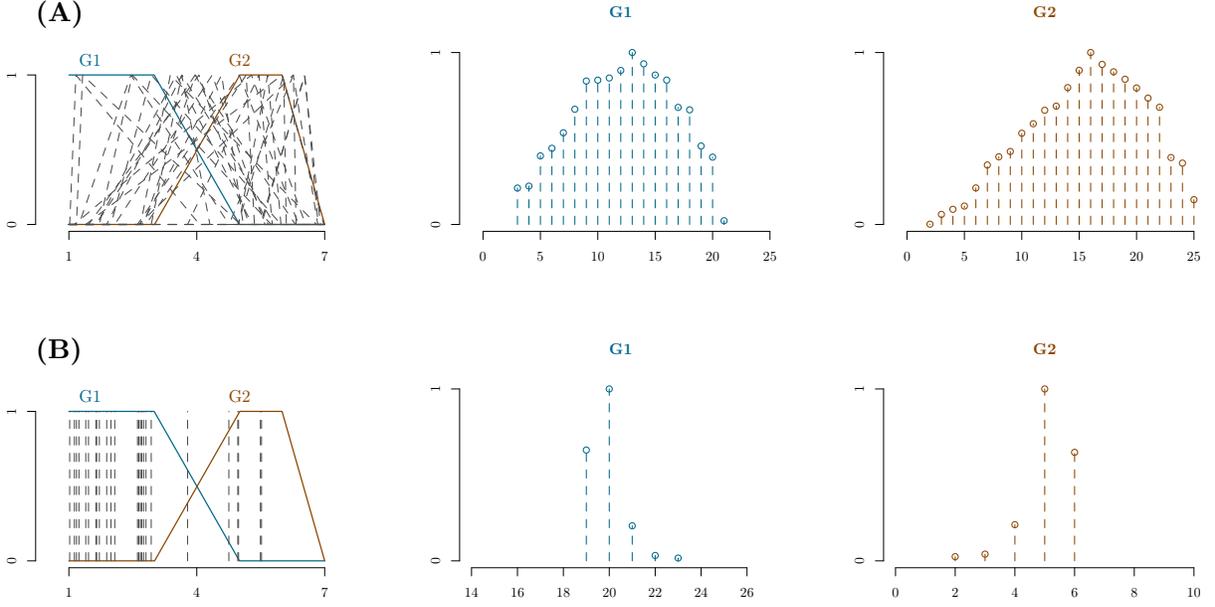}}
	\caption{Examples of fuzzy granules and fuzzy counts for (A) fuzzy triangular observations and fuzzy trapezoidal categories and (B) crisp observations and fuzzy trapezoidal categories. Note that in both cases frequencies are represented as generalized natural numbers.}
	\label{fig1}
\end{figure}

Finally, it is relevant to point out that Eqs. \eqref{eq2}-\eqref{eq3} are quite general and can be applied for the cases of fuzzy observations/fuzzy categories, crisp observations/fuzzy categories, and fuzzy observations/crisp categories. In this context, crisp observations and crisp categories can be realized by means of degenerated fuzzy sets and rectangular fuzzy sets, respectively. For the special case of crisp observations/crisp categories, the resulting fuzzy set $\fuzzysetij{n}{rc}$ is degenerate. Figure \ref{fig1} shows an exemplary case of fuzzy frequencies for fuzzy observations and fuzzy categories (Figure \ref{fig1}-A, middle and rightmost panels) and crisp observations and fuzzy categories as well (Figure \ref{fig1}-B, middle and rightmost panels). 

\begin{algorithm}
	\caption{Computing fuzzy frequencies}
	\label{alg1}
	\begin{algorithmic}
		\State
		\Procedure{Main}{$\mathbf{ \tilde x}_j,\mathbf{ \tilde x}_k,\mathbf{\tilde g}_j$, $\mathbf{\tilde g}_k$}
		\For{$r=1,\ldots,R \text{~\textbf{and}~} c=1,\ldots,C$} 
		\State $\boldsymbol{\epsilon}_r^j ~\gets \text{DoI}(\mathbf{\tilde x}_j,{\tilde g}_r)$ \Comment{\footnotesize Compute degrees of inclusion \normalsize}
		\State $\boldsymbol{\epsilon}_c^k ~\gets \text{DoI}(\mathbf{\tilde x}_k,{\tilde g}_c)$
		\State $\boldsymbol{\epsilon}^{jk}_{rc} \gets \min(\boldsymbol{\epsilon}_r^j,\boldsymbol{\epsilon}_c^k)$ \Comment{\footnotesize Compute joint degree of inclusion \normalsize}
		\State $\mathbf Z \gets (\boldsymbol{\epsilon}^{jk}_{rc} \mathbf 1_I^T - \mathbf 1_I^T\boldsymbol{\epsilon}^{jk}_{rc})$ 
		\For{$i=1,\ldots,I$} \State $z[i] \gets \mathbf 1_I^T \mathcall H(Z[:,i])$\EndFor 
		\For{$n=0,\ldots,I$} 
		\State $\mu_\text{\tiny FGC}[n] \gets ~\max( \mathcall H(\mathbf z-n)\odot \boldsymbol{\epsilon}_{rc}^{jk} )$ \Comment{\footnotesize Fuzzy counting functions \normalsize}
		\State $\mu_\text{\tiny FLC}[n] \gets ~1- \max( \mathcall H(\mathbf z-(n+1))\odot \boldsymbol{\epsilon}_{rc}^{jk} )$ 
		\State ~$\fuzzysetij{n}{rc}^{jk}[n] \gets ~\min ( \mu_\text{\tiny FLC}[n], \mu_\text{\tiny FGC}[n] )$ \Comment{\footnotesize Compute fuzzy frequencies \normalsize}
		\EndFor
		\EndFor
		\State \textbf{return} ~$\boldsymbol{\fuzzysetij{n}{}}^{jk}$
		\EndProcedure
		\Statex
		\Procedure{DoI}{$\mathbf{\tilde x},{\tilde g}$}
		\For{$i=1,\ldots,I$}
		\State ${\epsilon}[i] \gets  \int_{x} \min \big(\fuzzysetij{x}{[i]}(x),\fuzzysetij{g}{}(x)\big) dx ~\Big/ \max\big(1, \int_{x} \fuzzysetij{x}{[i]}(x) dx \big)$ \Comment{\footnotesize Ratio of fuzzy cardinality \normalsize}
		\EndFor
		\State \textbf{return} ~$\boldsymbol{\epsilon}$
		\EndProcedure
		\Statex\vspace{0.25cm}
		\begin{minipage}[t]{0.8\textwidth}
			\footnotesize \textsc{Note}: The algorithm requires as input the $I\times 1$ arrays of fuzzy observations $\mathbf{ \tilde x}_j$ and $\mathbf{ \tilde x}_k$ along with the fuzzy categories $\mathbf{ \tilde g}_j$ and $\mathbf{ \tilde g}_k$ for the $j,k$-th pair of variables and returns as output the $R\times C$ array of membership functions $\boldsymbol{\fuzzysetij{n}{}}^{jk} = (\boldsymbol{\fuzzysetij{n}{11}},\ldots,\boldsymbol{\fuzzysetij{n}{rc}},\ldots,\boldsymbol{\fuzzysetij{n}{RC}})$ associated to each fuzzy count $\tilde n_{rc}^{jk}$.\normalsize	
		\end{minipage}
	\end{algorithmic}
\end{algorithm}

\section{LLCs for fuzzy frequency tables}\label{sec:3}

In this section we describe the statistical procedure for computing latent linear correlations when observations are in the general form of fuzzy frequencies. 

\subsection{Model}\label{sec:3_1}

Let $X = (X^j_i,X^k_i) ~ i=1,\ldots,I$ be a collection of pairs of continuous random variables ($j,k \in \{1,\ldots,J\}, j\neq k$) following the bivariate Gaussian distribution centered at zero with correlation parameter $\rho_{jk} \in [-1,1]$ and density
\begin{equation}\label{eq4}
	f_X(\mathbf x;\rho_{jk}) = \frac{1}{2\pi\sqrt{1-\rho_{jk}^2}}\exp\left(-\frac{1}{2}\left[ \frac{(x^{j})^2+(x^{k})^2-2x^{j}x^k \rho_{jk}}{1-\rho_{jk}^2} \right] \right)
\end{equation}
for $-\infty < x^j < \infty$ and $-\infty < x^k < \infty$.
\noindent Without loss of generality, consider the collection of fuzzy observations $$\mathbf{\tilde y} = \{(\tilde y^{j}_1,\tilde y^{k}_1),\ldots,(\tilde y^{j}_i,\tilde y^{k}_i),\ldots,(\tilde y^{j}_I,\tilde y^{k}_I)\}$$ which relates to the (latent) bivariate Gaussian model in Eq. \eqref{eq4} via the constraint
\begin{equation}\label{eq4b}
	(\tilde y^j_i \in \tilde g_r^j) \land (\tilde y^k_i \in \tilde g_c^k) \quad\text{iff}\quad (X^j_i,X^k_i) \in (\tau^{X^j}_{r-1}, \tau^{X^j}_{r}] \times (\tau^{X^k}_{c-1}, \tau^{X^k}_{c}]\subset\mathbb R^2
\end{equation}
where $\in$ is intended as fuzzy membership, $(\tilde g_r^j,\tilde g_c^k)$ are observed fuzzy categories or granules, and the arrays $\boldsymbol{\tau}_{X^j} = (\tau^{X_j}_0,\ldots,\tau^{X_j}_r,\ldots,\tau^{X_j}_R)$ and $\boldsymbol{\tau}_{X^k} = (\tau^{X_k}_0,\ldots,\tau^{X_k}_c,\ldots,\tau^{X_k}_C)$ are \textit{thresholds} of the bivariate support $\mathbb R^2$ under the conventions $\tau^{X_j}_0 = \tau^{X_k}_0 = -\infty$ and $\tau^{X_j}_R = \tau^{X_k}_C = \infty$. Note that since fuzzy numbers encompass crisp observations and crisp categories as special cases (i.e., degenerated and rectangular fuzzy numbers, respectively), the expression \eqref{eq4b} can be used for the non fuzzy case as well. For instance, the simplest situation involving non fuzzy observations and non fuzzy categories can be obtained rewriting the left part of the constraint as $(\mathring y^j_i = r) \land (\mathring y^k_i = c)$, which indicates that crisp observations take the indices of the categories. 

The parameter space for the LLCs model is  
\begin{equation*}
	\boldsymbol{\theta} = \{\rho_{jk},\boldsymbol{\tau}_{X^j},\boldsymbol{\tau}_{X^k}\} \in [-1,1]\times \mathbb R^{R-1}\times \mathbb{R}^{C-1}
\end{equation*}
whereas the log-likelihood function takes the following form in the case of independent and identically distributed fuzzy observations \cite{olsson1979maximum,lee1988estimation}:
\begin{align}
	\ln \mathcall L(\boldsymbol{\theta};\mathbf{\widetilde N}) & = K - \sum_{r=1}^{R}\sum_{c=1}^{C} ~\sum_{n\in\mathbb N_0} n~\fuzzysetij{n}{rc}^{jk}(n) \ln \pi_{rc}^{jk}(\boldsymbol{\theta}) \nonumber\\
	& = K - \sum_{r=1}^{R}\sum_{c=1}^{C} ~\sum_{n\in\mathbb N_0} n~\fuzzysetij{n}{rc}^{jk}(n) \ln \int_{\tau_{r-1}^{X^j}}^{\tau_{r}^{X^j}} \int_{\tau_{c-1}^{X^k}}^{\tau_{c}^{X^k}} f_X(\mathbf x;\rho_{jk}) ~dx^j dx^k \label{eq5}
\end{align}
where $f_X(\mathbf x;\rho_{jk})$ is the model's density in Eq. \eqref{eq4}, $\fuzzysetij{n}{rc}^{jk}(n)$ is the $rc$-th fuzzy count, and $K$ is a constant term. Note that $f_X(\mathbf x;\rho_{jk})$ is not fuzzy in this context and its realizations represent unobserved (latent) quantities. The evaluation of $(\tilde y^j_i \in \tilde g_r^j) \land (\tilde y^k_i \in \tilde g_c^k)$ gives raise to a collection of fuzzy counts $\tilde n_{11}^{jk},\ldots,\tilde n_{rc}^{jk},\ldots,\tilde n_{RC}^{jk}$ acting as \textit{possibilistic constraints} on the unobserved non-fuzzy counts which would be observed if fuzziness was missed. As such, the expression $\fuzzysetij{n}{rc}^{jk}(n_{rc}) \in [0,1]$ should be interpreted as the possibility that the crisp count $n_{rc}$ has to occur, with $\fuzzysetij{n}{rc}^{jk}( n_{rc} ) = 1$ indicating that $n_{rc}$ is fully possible. According to the epistemic viewpoint on fuzzy statistics \cite{couso2014statistical}, the sampling process is thought as being the consequence of a two-stage generation mechanism, the first of which is a random experiment and the second is a non-random fuzzification of the outcome being realized. As an example of this schema, consider the simplest case of crisp observations (e.g., {income} and {tobacco use}) that are classified by a group of raters or an automatic classification system on the basis of fuzzy categories (e.g., income levels: low, medium, high; tobacco use: none, sporadic, habitual). Stated in this way, the fuzzy frequencies associated to income and tobacco use encapsulate two sources of uncertainty, namely the \textit{random component} due to the sampling process and the \textit{non-random component} due to the post-sampling fuzzy classification. 

\subsection{Parameter estimation}\label{sec:3_2}

To estimate $\boldsymbol{\theta}$ we adopt the Olsson's two-stage approach for latent linear correlations which iteratively alternates between approximating $\boldsymbol{\hat\tau}$ from the observed count data and maximizing Eq. \eqref{eq5} with respect to $\hat\rho$ given the current thresholds \cite{olsson1979maximum}. In the case of fuzzy data, this procedure can be implemented using a variant of the Expectation-Maximization algorithm generalized to the case of fuzzy observations \cite{denoeux2011maximum}. Likewise for the standard EM algorithm, the fuzzy-EM version alternates between the E-step, which requires computing the expected complete log-likelihood given the candidate $\boldsymbol{\theta}' = \boldsymbol{\theta}^{(q-1)}$, and the M-step, which maximizes the expected complete log-likelihood w.r.t. $\boldsymbol{\theta}^{(q)}$. More precisely, in the fuzzy-EM algorithm the complete-data log-likelihood is that obtained if the matrix of counts $\mathbf N_{R\times C}^{jk}$ was precisely observed, namely:
\begin{equation}\label{eq6}
	\ln \mathcall L(\boldsymbol{\theta};\mathbf{ N}) = \ln I! - \sum_{r=1}^{R}\sum_{c=1}^{C} {n}_{rc}^{jk} \ln \pi_{rc}^{jk}(\boldsymbol{\theta}) - \sum_{r=1}^{R}\sum_{c=1}^{C} \ln n_{rc}^{jk}!
\end{equation}
Given the estimates $\boldsymbol{\theta}'$, the E-step for the $(q)$-th iteration consists of computing the $\mathcall Q$-function via conditional expectation on the observed fuzzy counts:
\begin{align}
	\mathcall Q(\boldsymbol{\theta},\boldsymbol{\theta}') &= \Expp{\boldsymbol{\theta}'}{\ln\mathcall L(\boldsymbol{\theta};\mathbf{N}) \Big| \mathbf{\widetilde{N}}}\nonumber\\
	& \propto \sum_{r=1}^{R}\sum_{c=1}^{C} ~\Expp{\boldsymbol{\theta}'}{{N}_{rc}^{jk}\Big| {\tilde{n}_{rc}}^{jk}} \ln \pi_{rc}^{jk}(\boldsymbol{\theta}) - 
	\Expp{\boldsymbol{\theta}'}{ \ln N_{rc}^{jk}! \Big| {\tilde{n}_{rc}}^{jk}} \label{eq7}
\end{align}
The conditional expectations involve the density of a discrete random variable $N_{rc}$ conditioned on a fuzzy event $\tilde{n}_{rc}$ that, under the multinomial schema for random counts, can reasonably be modeled as Binomial \cite{agresti2003categorical}. Thus, using the definition of fuzzy probability, $N_{rc}|\tilde{n}_{rc}$ is as follows:
\begin{align}
	&p_{N_{rc}^{jk}|\tilde{n}_{rc}^{jk}}(n;\pi_{rc}^{jk}(\boldsymbol{\theta})) = \frac{\Probb{N_{rc}^{jk},\tilde{n}_{rc}^{jk}}{\boldsymbol{\theta}}}{\Probb{\tilde{n}_{rc}^{jk}}{\boldsymbol{\theta}}} = \frac{\fuzzysetij{n}{jk}^{jk}(n) p_{N_{rc}^{jk}}(n;\pi_{rc}^{jk}(\boldsymbol{\theta}))}{\sum_{n\in\mathbb N_0} n~\fuzzysetij{n}{jk}^{jk}(n) p_{N_{rc}^{jk}}(n;\pi_{rc}^{jk}(\boldsymbol{\theta}))} \label{eq8}\\
	& \pi_{rc}^{jk}(\boldsymbol{\theta}) = \int_{\tau_{r-1}^{X^j}}^{\tau_{r}^{X^j}} \int_{\tau_{c-1}^{X^k}}^{\tau_{c}^{X^k}} f_X(\mathbf x;\rho_{jk}) ~dx^j dx^k \label{eq8b}
\end{align}
where $p_{N_{rc}^{jk}}= \mathcall Bin(n;\pi_{rc}^{jk}(\boldsymbol{\theta}))$ and $f_X(\mathbf x;\rho_{jk})$ is the latent model's density in Eq. \eqref{eq4}. Note that the quantity $I\pi_{rc}^{jk}(\boldsymbol{\theta})$ is the reconstructed count from the bivariate latent model given the current parameters $\boldsymbol{\theta}'$ \cite{shiina2017polychoric}. The linear form of the expectations in Eq. \eqref{eq7} is:
\begin{equation}\label{eq9a}
	\Expp{\boldsymbol{\theta}'}{N_{rc}^{jk}\Big| {\tilde{n}_{rc}}^{jk}} = \sum_{n\in\mathbb N_0} n ~p_{N_{rc}^{jk}|\tilde{n}_{rc}^{jk}}(n;\pi_{rc}^{jk}(\boldsymbol{\theta}')) 
\end{equation}
whereas, since it is not involved in the M-step of the algorithm, the non linear expectation is provided in Appendix \ref{apx1} for the sake of completeness.

Finally, the M-step for the $(q)$-th iteration requires maximizing the functional $\mathcall Q(\boldsymbol{\theta},\boldsymbol{\theta}')$ with respect to $\boldsymbol{\theta}$. Given the filtered counts at the current iteration $\widehat{\mathbf N}^{jk}_{R\times C}$ (see Eq. \ref{eq9a}), the Olsson's two-stage estimation approach requires the estimation of thresholds from the cumulative marginals of filtered counts first:
\begin{align}
	& \widehat{\boldsymbol{\tau}}_{X^j}^{(q)} = \Phi^{-1}\left( \mathbf A_{R\times R}\widehat{\mathbf N}^{jk}\mathbf{1}_{C} \right) \label{eq10a}\\
	& \widehat{\boldsymbol{\tau}}_{X^k}^{(q)} = \Phi^{-1}\left( \mathbf A_{C\times C}(\widehat{\mathbf N}^{jk})^T\mathbf{1}_{R} \right) \label{eq10b}
\end{align}
where $\mathbf A$ is a lower triangular matrix of ones, $\mathbf 1$ is a vector of appropriate order of all ${1}/{I}$, and $\Phi$ is the Gaussian univariate distribution function with mean zero and unitary variance. Next, conditioned on $\{\widehat{\boldsymbol{\tau}}_{X^j}^{(q)},\widehat{\boldsymbol{\tau}}_{X^k}^{(q)}\}$, the remaining parameter is found by solving the score equation of $\mathcall Q(\boldsymbol{\theta},\boldsymbol{\theta}^{(q)})$ numerically w.r.t. $\rho_{jk}$:
\begin{equation}\label{eq10c}
	\mathcall U_{\rho_{jk}} = \frac{\partial \mathcall Q\left(\rho_{jk},\{\widehat{\boldsymbol{\tau}}_{X^j}^{(q)},\widehat{\boldsymbol{\tau}}_{X^k}^{(q)}\}\right)}{\partial {\boldsymbol\pi^{jk}}} \frac{\partial \boldsymbol\pi^{jk}}{\partial \rho_{jk}} = 0 
\end{equation}
The algorithm proceeds iteratively until the log-likelihood does not increase significantly. Table \ref{tab1} summarizes the fuzzy-EM algorithm for the LLCs model.

\begin{table}[!h]
	\centering
	\begin{tabular}{llr} \toprule[0.03cm]
		\small{\textbf{Algorithm 2}} &{Olsson's two-stage approach via fuzzy-EM algorithm}&\\ \hline
		&&\\[-0.20cm]
		&for $j\in (1,\ldots,J)$ and $k\in (1,\ldots,J)$, $j\neq k$, \underline{do}: &\\[0.3cm]		
		$\mathbf{q=1:}$ & \small{Set} ~\normalsize$\boldsymbol\theta^{(q)} = (\rho_{jk}^0,\boldsymbol{\tau}_{X^j}^0,\boldsymbol{\tau}_{X^k}^0)$, $~l^{(q)} = l^0$, $~\epsilon=1e^{-09}$ & \textsc{initialization}\\[0.4cm]
		$\mathbf{q>1:}$ & \small{Compute} ~\normalsize $\boldsymbol{\pi}^{jk}(\boldsymbol{\theta}^{(q-1)})$ \hspace{0.3cm}\small{from Eq. \eqref{eq8b}}\normalsize & \textsc{E-Step}\\
		& \small{Compute} ~\normalsize $\widehat{\mathbf N}^{jk}$ \hspace{0.3cm} given $\boldsymbol{\theta}^{(q-1)}$  \hspace{0.3cm}\small{from Eq. \eqref{eq9a}}\normalsize &\\
		& \small{Compute} ~\normalsize $\ln \widehat{\mathbf N}^{jk}!$ given $\boldsymbol{\theta}^{(q-1)}$  \hspace{0.3cm}\small{from Eq. \eqref{eqA1}}\normalsize &\\[0.5cm]
        & \small{Compute} ~\normalsize $\{\widehat{\boldsymbol{\tau}}_{X^j}^{(q)}, \widehat{\boldsymbol{\tau}}_{X^k}^{(q)}\}$ \hspace{0.3cm}\small{from Eqs. \eqref{eq10a}-\eqref{eq10b}}\normalsize & \textsc{M-Step}\\[0.15cm]
        & \small{Set} ~ \normalsize $\boldsymbol\theta^{(q)} = \left(\rho_{jk}^{(q-1)},\boldsymbol{\tau}_{X^j}^{(q)},\boldsymbol{\tau}_{X^k}^{(q)}\right)$ & \\[0.15cm]
        & \small{Solve} ~\normalsize $\frac{\partial}{\partial \rho_{jk}}\mathcall Q(\boldsymbol{\theta},\boldsymbol{\theta}^{(q)}) = 0$ w.r.t. $\rho_{jk}$ \hspace{0.3cm}\small{see Eq. \eqref{eq10c}}\normalsize &\\[0.15cm]
        & \small{Set} ~ \normalsize $\boldsymbol\theta^{(q)} = \left(\rho_{jk}^{(q)},\boldsymbol{\tau}_{X^j}^{(q)},\boldsymbol{\tau}_{X^k}^{(q)}\right)$ & \\[0.5cm]
        & \small{Evaluate} ~\normalsize $l^{(q)} = \ln \mathcall L(\boldsymbol{\theta}^{(q)};\mathbf{\widehat N})$ \hspace{0.3cm}\small{see Eq. \eqref{eq6}}\normalsize & \textsc{Finalization}\\
        & \small{Compute} ~\normalsize $l_\delta = (l^{(q)}-l^{(q-1)})$ &\\
        & {If} $~l_\delta < \epsilon$, \small{set} ~\normalsize $\widehat{\boldsymbol\theta} = \boldsymbol\theta^{(q)}$ {and \underline{stop} the algorithm} & \\
        & $\mathbf R[j,k]=\rho^{(q)}_{jk}$ &\\
        &&\\ \hline 
	\end{tabular}
	\caption{Expectation-Maximization algorithm for estimating $\boldsymbol{\theta} = (\boldsymbol{\tau}_{X^j},\boldsymbol{\tau}_{X^k},\rho_{jk})$ in LLCs model with fuzzy frequency data.}
	\label{tab1}
\end{table}

\subsection{Remarks}\label{sec:3_3}

\noindent \textit{About the convergence of the algorithm}. Given a candidate $\boldsymbol{\theta}'$, the fuzzy-EM starts by constructing the surrogate $\mathcall Q(\boldsymbol{\theta},\boldsymbol{\theta}')$ that lower bounds the observed data log-likelihood $\ln \mathcall L(\boldsymbol{\theta};\mathbf{\widetilde{N}})$ (E-step). Next, it is maximized to get the current estimates $\boldsymbol{\theta}^{(q)}$ (M-step), which is in turn used to construct a new lower bound $\mathcall Q(\boldsymbol{\theta},\boldsymbol{\theta}^{(q)})$ in the next iteration to get a new estimate $\boldsymbol{\theta}^{(q+1)}$. The estimates in the M-step are chosen so that $\mathcall Q(\boldsymbol{\theta},\boldsymbol{\theta}^{(q)}) \geq \mathcall Q(\boldsymbol{\theta},\boldsymbol{\theta}')$, which forms the base of the monotonicity condition $\ln \mathcall L(\boldsymbol{\theta}^{(q+1)};\mathbf{\widetilde{N}}) \geq \ln \mathcall L(\boldsymbol{\theta}^{(q)};\mathbf{\widetilde{N}})$ \cite{mclachlan2007algorithm}. As for the standard case, the monotonicity of the sequence $\{\ln \mathcall L(\boldsymbol{\theta}^{(q)}\}_{q\in\mathbb{N}}$ implies the convergence to a stationary value, which can be global or local depending on the characteristics of the log-likelihood function and the starting point $\boldsymbol{\theta}^0$. A sketch of the proof of the monotonicity of the fuzzy-EM for the LLCs is provided in Appendix \ref{apx2} whereas the formal equivalence between EM and fuzzy-EM is detailed in \cite{denoeux2011maximum,quost2016clustering}.\\

\noindent \textit{About the starting values of the algorithm}. Suitable starting values $\boldsymbol{\theta}^0$ can be obtained by first defuzzifying the observed fuzzy frequencies matrix $\widetilde{\mathbf N}^{jk}$ to obtain non fuzzy counts and then applying the standard Olsson's two-stage approach \cite{olsson1979maximum} on defuzzified data. In general, this yields to convenient starting values. In the LLCs model, defuzzification can be performed via \textit{mean} or \textit{max}-based procedures as follows: $\hat n_{rc}^{\text{\tiny mean}} \approxx \sum_{n\in\mathbb N_0} n \fuzzysetij{n}{rc}(n) / \left(\sum_{n\in\mathbb N_0} \fuzzysetij{n}{rc}(n) ~ \right) $, $\hat n_{rc}^{\text{\tiny max}} = \max \{n\in\mathbb{N}_0: \fuzzysetij{n}{rc}(n) = \max_{z \in \mathbb{N}_0}~ \fuzzysetij{n}{rc}(z) \}$, $~r=1,\ldots,R$, $~c=1,\ldots,C$. \\

\noindent \textit{About the term $p_{N_{rc}|\tilde{n}_{rc}}(n;\pi_{rc}(\boldsymbol{\theta}))$}. The term $p_{N_{rc}|\tilde{n}_{rc}}$ represents the density of a non fuzzy random variable conditioned on fuzzy numbers and can mathematically be interpreted as the combination of two independent components, namely the \textit{random} mechanism underlying the sampling process and the observer's partial knowledge (\textit{imprecision}) about the sample realizations. In this sense, as it weights each fuzzy datum by the probability that it has to occur \cite{zah1968probability}, $p_{N_{rc}|\tilde{n}_{rc}}$ should not be confused with the mean-based defuzzification of fuzzy numbers. A nice property of this formulation is that fuzziness vanishes when precise observations are available. Indeed, the conditional density involving a degenerated fuzzy number $\mathring{n}_{rc}$ boils down to a degenerated discrete density $p_{N_{rc}|\mathring{n}_{rc}}$ with nonzero probability masses only for those $n$ such that $\fuzzysetij{n}{rc}(n)=1$. As a consequence, the fuzzy-EM procedure reduces to standard Olsson's two-stage maximum likelihood estimation. In general, there are a number of ways for plugging-in non-stochastic components of uncertainty into $p_{N_{rc}|\tilde{n}_{rc}}$, such as those involving imprecise probability \cite{augustin2014introduction}, conditional probability \cite{coletti2004conditional}, belief measures \cite{yager1982generalized}, and random fuzzy variables \cite{gil2006overview}. \\

\noindent \textit{About the computation of standard errors and inference}. Standard errors for $\hat\rho_{jk}$ can be computed as a byproduct of the EM procedure \cite{mclachlan2007algorithm}. In particular, the following approximation of the information matrix is required $\mathcall I_{\hat\rho_{jk}} \approxx \mathcall I_{\hat\rho_{jk}}^e = \sum_{r=1}^R\sum_{c=1}^C \mathcall{U}_{\hat\rho_{jk}}^{(rc)}\mathcall{U}_{\hat\rho_{jk}}^{(rc)}$, with $\mathcall{U}_{\hat\rho_{jk}}^{(rc)}$ being the score functional for the $rc$-th observation calculated at $\hat{\rho}_{jk}$, to get the standard error $\sigma_{\hat\rho_{jk}} = ({1/\mathcall I_{\hat\rho_{jk}}^e})^{\frac{1}{2}}$. Alternatively, they can also be obtained by means of non-parametric or parametric bootstrap techniques \cite{zhang2006bootstrap}. Finally, it should be remarked that inference about $\rho_{jk}$ can be made based on the asymptotic results of fuzzy likelihood ratio statistics (e.g., see \cite{berkachy2019fuzzy}).\\

\noindent \textit{About the polychoric correlation matrix $\mathbf{R}_{J\times J}$}. As for the standard approach used in computing polychoric correlation matrices (e.g., see: \cite{olsson1979maximum,joreskog1994estimation}), also in the case of fuzzy data the matrix of latent linear correlations is obtained by calculating each element $\rho_{jk}$ of the correlation matrix pairwise. Although this approach offers a simple and effective alternative to more challenging methods (e.g., see: \cite{lee1987two,song2003full}), in some circumstances it may lead to non-positive definite correlation matrices. This can be problematic, especially when such matrices are used as input of other statistical models such as factor analyses or SEMs \cite{lorenzo2021not}. In these cases, eigenvalue decomposition based smoothing \cite{knol1989least}, least squares \cite{knol1989least} or Dykstra's \cite{higham2002computing} corrections constitute workable solutions to solve this issue.

\section{Simulation study}\label{sec:4}

The aim of this simulation study is twofold. First, we wish to evaluate the performances of fuzzy-EM algorithm in estimating parameters of the LLCs model and, second, to investigate whether the standard Olsson's maximum likelihood procedure performs as good as the proposed method if applied on max-based and mean-based defuzzified data. The case $J=2$ has been considered for the sake of simplicity. The Monte Carlo study has been performed on a (remote) HPC machine based on 16 cpu Intel Xeon CPU E5-2630L v3 1.80 GHz,16x4 GB Ram whereas computations and analyses have been done in the \texttt{R} framework for statistical analyses.\\

\noindent \textit{Design}. The design of the study involved three factors, namely (i) $I\in\{150,250,500,1000\}$, (ii) $\rho^0 \in \{0.15,0.50,0.85\}$, (iii) $R=C\in\{4,6\}$, which were varied in a complete factorial design with $4\times 3\times 2=24$ possible combinations. The threshold parameters were held fixed under the equidistance hypothesis \cite{joreskog1994estimation}, namely $\boldsymbol{\tau}^0_{X^j}=\boldsymbol{\tau}^0_{X^k}=(-2.00,-0.66,0.66,2.00)$ for the conditions with $R=C=4$ and $\boldsymbol{\tau}^0_{X^j}=\boldsymbol{\tau}^0_{X^k}=(-2.00,-1.20,-0.40,0.40,1.20,2.00)$ for $R=C=6$. For each combination, $B=5000$ samples were generated yielding to $5000\times 24 = 120000$ new data and an equivalent number of parameters.\\

\noindent \textit{Data generation and procedure}. Let $I_{a}$, $\rho^0_{b}$, $R_{d}=C_d$ be distinct levels of the factors $I$, $\rho^0$, $R$, and $C$. Then, fuzzy frequency data have been generated according to the following procedure. For each $r=1,\ldots,R_d$ and $c=1,\ldots,C_d$:
\begin{enumerate}
	\item[(i)] Set $n_{rc} = I_a\pi_{rc}$ (see Eq. \eqref{eq8b}) given $\boldsymbol{\tau}^0_{X^j}$, $\boldsymbol{\tau}^0_{X^k}$, $\rho^0_{b}$, and $I_a$
	\item[(ii)] the imprecision concerning $n_{rc}$ was generated as follows: $m_1\sim \mathcall{G}amma_{\text{d}}(\alpha_{m_1},\beta_{m_1})$ where $\alpha_{m_1}= 1+n_{rc}\beta_{m_1}$, $\beta_{m_1}=(n_{rc}+n_{rc}^2+4s^2_1)^{\frac{1}{2}} \big/ 2s^2_1$, $s_1 \sim \mathcall{G}amma_{\text{d}}(\alpha_{s_1},\beta_{s_1})$, $\alpha_{s_1}= 1+m_0\beta_{s_1}$, $\beta_{s_1}=(m_0+m_0^2+4s^2_0)^{\frac{1}{2}} \big/ 2s^2_0$, $m_0=1$ and $s_0=0.25$, with $\mathcall{G}amma_{\text{d}}$ indicating the discrete Gamma random variable with shape and rate being reparameterized in terms of mean $m$ and variance $s$
	\item[(iii)] the fuzzy set associated to $\tilde n_{rc}$ was obtained via the following probability-possibility transformation: $\boldsymbol\xi_{\tilde n_{rc}} = f_{\mathcall{G}_{\text{d}}}(\mathbf n;\alpha_{rc},\beta_{rc}) \big / \max f_{\mathcall{G}_{\text{d}}}(\mathbf n;\alpha_{rc},\beta_{rc})$, with $\mathbf n=\{0,1,\ldots,I_{a}\}$, $\alpha_{rc} = 1+m_1\beta_{s_1}$, $\beta_{s_1}=1+(m_1+m_1^2+4s_1^2)^{\frac{1}{2}} / 2s_1^2 $, $\beta_{rc}=(m_1+m_1^2+4s^2_1)^{\frac{1}{2}} \big/ 2s^2_1$, and $f_{\mathcall{G}_{\text{d}}}(n;\alpha_{rc},\beta_{rc})$ being the discrete Gamma density normalized to one in order to mimics the behavior of a normal fuzzy set \cite{dubois2012fundamentals}. The discrete density $f_{\mathcall{G}_{\text{d}}}$ is computed as difference of survival functions of the continuous Gamma density $S_{\mathcall{G}}(x) - S_{\mathcall{G}}(x+1)$ \cite{chakraborty2012discrete,extraDistr20}.
\end{enumerate} 
Note that step (ii) is required in order to make crisp counts entirely imprecise so that $\tilde n_{rc}$ is no longer centered on $n_{rc}$. Finally, parameters $\boldsymbol{\theta} = \{\rho,\boldsymbol{\tau}_{X^j},\boldsymbol{\tau}_{X^k}\}$ were estimated from the fuzzy counts $\mathbf{\widetilde N}_{R_d\times C_d}$ using the fuzzy-EM algorithm (fEM) and the standard Olsson's two-stage maximum likelihood on max-based (dML-max) and mean-based (dML-mean) defuzzified counts (see Sect. \ref{sec:3_3}). \\

\noindent \textit{Outcome measures}. For each condition of the simulation design, the three methods (i.e., fEM, ML-max, ML-mean) were evaluated in terms of \textit{bias} and \textit{root mean square errors}. In addition, for each method thresholds were aggregated to form a scalar statistic, namely $\widehat{\boldsymbol{\tau}} = \mathbf 1_{R_d}^T\widehat{\boldsymbol{\tau}}_{X^j}$ and $\widehat{\boldsymbol{\tau}} = \mathbf 1_{C_d}^T\widehat{\boldsymbol{\tau}}_{X^k}$ (note that $\boldsymbol{\tau}_{X^j}$ and $\boldsymbol{\tau}_{X^k}$ are equal by design). \\

\noindent \textit{Results}. Tables \ref{tab2a}-\ref{tab2d} report the results of the simulation study with regards to $\hat\rho$ and $\hat{\boldsymbol{\tau}}$ for both $R=C=4$ and $R=C=6$ cases. We begin with the correlation parameter $\rho$ for the case $R=C=4$ (see Table \ref{tab2a}). Considering $\rho^0=0.15$, the methods showed negligible bias in estimating $\rho$. However, they differed in terms of RMSE, with fEM showing lower values with increasing sample size if compared to dML-max and dML-mean. With increasing correlation length ($\rho^0>0.15$), bias of estimates as well as RMSE were more pronounced for dML-max and dML-mean. The same results were also observed for the case with $R=C=6$ (see Table \ref{tab2b}). With regards to the overall statistic $\hat{\boldsymbol{\tau}}$ for the threshold parameters, all the methods achieved comparable results regardless of $\rho^0$. In particular, fEM showed slightly higher bias and RMSE then dML-max and dML-mean methods across $R=C=4$ (see Table \ref{tab2c}) and $R=C=6$ (see Table \ref{tab2d}) conditions. To further investigate these results, we studied average bias and variance of estimates for $\boldsymbol{\hat\tau}_{X^j}$ (or $\boldsymbol{\hat\tau}_{X^k}$) as a function of sample size $I$ and $\rho^0$. We found that the leftmost and rightmost thresholds tended to be slightly larger for fEM as opposed to the innermost thresholds for both $R=C=4$ (see Supplementary Materials, Figure S1) and $R=C=6$ conditions (see Supplementary Materials, Figure S2). Moreover, the variance of estimates for the leftmost and rightmost thresholds was higher if compared to the innermost thresholds (see Supplementary Materials, Table S2) but, as expected, it reduced with increasing sample size regardless of $\rho^0$. This is not surprising given that we implemented a standard LLCs model in which no particular constraints were applied on threshold estimates, such as $\mathbf 1_{R_d}^T\hat{\boldsymbol{\tau}}_{X^j} = 0$ (e.g., see \cite{foldnes2020pernicious}).\footnote{It should be remarked that the unconstrained approach is most common in LLCs-based applications, especially when the primary interest lies in making inference about $\rho$. In this case, the threshold parameters play an auxiliary role as they only affect the scale of the latent variables underlying the LLCs model (e.g., see \cite{lee2001maximum}). } Most importantly, according to the Gaussianity assumption underlying the LLCs model, estimated thresholds were symmetric and equidistant with respect to the fixed point zero (see Supplementary Materials, Table S1). Overall, the results suggest that fEM should be preferred over defuzzified maximum likelihood when the interest is in estimating the latent linear association $\rho$ among pairs of variables and fuzzy frequency statistics are available. On the contrary, for those particular cases where $\rho$ is known and the interest is in estimating the true threshold parameters, then standard Olsson's maximum likelihood method can directly be applied after defuzzifiyng observed fuzzy frequency counts.

\begin{table}[h!]
	\centering
	\begin{tabular}{lccccccc}
		\toprule
		\multirow{2}{*}{} &
		\multicolumn{2}{c}{fEM} &
		\multicolumn{2}{c}{dML-max} &
		\multicolumn{2}{c}{dML-mean} \\ \cmidrule(lr){2-3} \cmidrule(lr){4-5} \cmidrule(lr){6-7} 
		$R=C=4$& {\textit{bias}}& {\textit{rmse}}& {\textit{bias}}& {\textit{rmse}}& {\textit{bias}}& {\textit{rmse}}\\
		\midrule
		$\rho=0.15$ &&&&&&&\\[0.1cm]
		$I=150$ & 0.03401 & 0.08911 & -0.01653 & 0.11826 & -0.04354 & 0.08824 \\ 
		$I=250$ & 0.00455 & 0.05062 & -0.02821 & 0.08106 & -0.04020 & 0.06766 \\ 
		$I=500$ & 0.01047 & 0.02974 & 0.00311 & 0.04180 & -0.00743 & 0.03339 \\ 
		$I=1000$ & -0.00321 & 0.01515 & -0.01012 & 0.02441 & -0.01421 & 0.02276 \\[0.2cm]
		$\rho=0.50$ &&&&&&&\\[0.1cm] 
		$I=150$ & 0.01265 & 0.07236 & -0.08807 & 0.15014 & -0.17694 & 0.19253 \\ 
		$I=250$ & -0.03699 & 0.06349 & -0.12376 & 0.15052 & -0.17174 & 0.18119 \\ 
		$I=500$ & -0.00151 & 0.02688 & -0.04673 & 0.06983 & -0.08356 & 0.09120 \\ 
		$I=1000$ & -0.00050 & 0.01389 & -0.02226 & 0.03582 & -0.03921 & 0.04459 \\[0.2cm]
		$\rho=0.85$ &&&&&&&\\[0.1cm] 
		$I=150$ & 0.00194 & 0.04504 & -0.21865 & 0.25598 & -0.32889 & 0.33729 \\ 
		$I=250$ & -0.00285 & 0.02903 & -0.17042 & 0.19816 & -0.25843 & 0.26540 \\ 
		$I=500$ & -0.00104 & 0.01586 & -0.10519 & 0.12382 & -0.16418 & 0.16884 \\ 
		$I=1000$ & -0.00056 & 0.00810 & -0.06598 & 0.07880 & -0.10451 & 0.10760 \\
		\hline\bottomrule
	\end{tabular}
	\caption{Simulation study: Average bias and root mean square errors for $\rho$ in the condition $R=C=4$. Note that fEM is the fuzzy-EM algorithm whereas dML-max and dML-mean denote the standard maximum likelihood based on max-based and mean-based defuzzified counts.} 
	\label{tab2a}
\end{table}

\begin{table}[h!]
	\centering
	\begin{tabular}{lccccccc}
		\toprule
		\multirow{2}{*}{} &
		\multicolumn{2}{c}{fEM} &
		\multicolumn{2}{c}{dML-max} &
		\multicolumn{2}{c}{dML-mean} \\ \cmidrule(lr){2-3} \cmidrule(lr){4-5} \cmidrule(lr){6-7} 
		$R=C=6$& {\textit{bias}}& {\textit{rmse}}& {\textit{bias}}& {\textit{rmse}}& {\textit{bias}}& {\textit{rmse}}\\
		\midrule
		$\rho=0.15$ &&&&&&&\\[0.1cm] 
		$I=150$ & 0.02884 & 0.10022 & -0.01490 & 0.10067 & -0.04919 & 0.08355 \\ 
		$I=250$ & 0.00860 & 0.05720 & -0.02461 & 0.07169 & -0.04289 & 0.06501 \\ 
		$I=500$ & -0.01619 & 0.03395 & -0.02606 & 0.04555 & -0.03853 & 0.04869 \\ 
		$I=1000$ & 0.00064 & 0.01474 & -0.00539 & 0.01947 & -0.01021 & 0.01892 \\[0.2cm]
		$\rho=0.50$ &&&&&&&\\[0.1cm] 
		$I=150$ & -0.05241 & 0.10238 & -0.17183 & 0.20107 & -0.25228 & 0.26146 \\ 
		$I=250$ & -0.00259 & 0.04950 & -0.09811 & 0.12241 & -0.16604 & 0.17374 \\ 
		$I=500$ & -0.00644 & 0.02414 & -0.05111 & 0.06359 & -0.08845 & 0.09304 \\ 
		$I=1000$ & -0.00214 & 0.01222 & -0.02415 & 0.03205 & -0.04278 & 0.04594 \\[0.2cm]
		$\rho=0.85$ &&&&&&&\\[0.1cm] 
		$I=150$ & -0.00245 & 0.05111 & -0.24268 & 0.26670 & -0.38252 & 0.38855 \\ 
		$I=250$ & -0.01614 & 0.03414 & -0.18437 & 0.19946 & -0.28388 & 0.28831 \\ 
		$I=500$ & 0.00078 & 0.01412 & -0.09831 & 0.10892 & -0.16062 & 0.16358 \\ 
		$I=1000$ & -0.00219 & 0.00694 & -0.05869 & 0.06422 & -0.09167 & 0.09336 \\ 
		\hline\bottomrule
	\end{tabular}
	\caption{Simulation study: Average bias and root mean square errors for $\rho$ in the condition $R=C=6$. Note that fEM is the fuzzy-EM algorithm whereas dML-max and dML-mean denote the standard maximum likelihood based on max-based and mean-based defuzzified counts.} 
	\label{tab2b}
\end{table}

\begin{table}[h!]
	\centering
	\begin{tabular}{lccccccc}
		\toprule
		\multirow{2}{*}{} &
		\multicolumn{2}{c}{fEM} &
		\multicolumn{2}{c}{dML-max} &
		\multicolumn{2}{c}{dML-mean} \\ \cmidrule(lr){2-3} \cmidrule(lr){4-5} \cmidrule(lr){6-7} 
		$R=C=4$& {\textit{bias}}& {\textit{rmse}}& {\textit{bias}}& {\textit{rmse}}& {\textit{bias}}& {\textit{rmse}}\\
		\midrule
		$\rho=0.15$ &&&&&&&\\[0.1cm]
		$I=150$ & 0.11082 & 0.30335 & 0.03419 & 0.16958 & 0.11037 & 0.14607 \\ 
		$I=250$ & 0.07881 & 0.12035 & 0.01038 & 0.10643 & 0.03761 & 0.08513 \\ 
		$I=500$ & 0.02189 & 0.04664 & 0.01363 & 0.05270 & 0.01595 & 0.04706 \\ 
		$I=1000$ & 0.02916 & 0.03685 & 0.01158 & 0.02967 & 0.01314 & 0.02781 \\[0.2cm]
		$\rho=0.50$ &&&&&&&\\[0.1cm] 
		$I=150$ & 0.00597 & 0.12056 & 0.09273 & 0.16206 & 0.14804 & 0.17656 \\ 
		$I=250$ & 0.04446 & 0.09646 & 0.02888 & 0.10059 & 0.05722 & 0.09249 \\ 
		$I=500$ & 0.04445 & 0.06225 & 0.00718 & 0.05440 & 0.00284 & 0.04488 \\ 
		$I=1000$ & 0.01771 & 0.02797 & 0.00017 & 0.02606 & 0.00288 & 0.02370 \\[0.2cm]
		$\rho=0.85$ &&&&&&&\\[0.1cm] 
		$I=150$ & 0.06325 & 0.32381 & 0.08313 & 0.16289 & 0.13419 & 0.16534 \\ 
		$I=250$ & 0.03843 & 0.09054 & 0.03940 & 0.10066 & 0.06298 & 0.09680 \\ 
		$I=500$ & 0.01918 & 0.04347 & 0.02264 & 0.05314 & 0.02680 & 0.04995 \\ 
		$I=1000$ & 0.03036 & 0.03668 & 0.00637 & 0.02696 & 0.00632 & 0.02414 \\ 
		\hline\bottomrule
	\end{tabular}
	\caption{Simulation study: Average bias and root mean square errors for the aggregated thresholds $\widehat{\boldsymbol{\tau}} = \mathbf 1_{R_d}^T\widehat{\boldsymbol{\tau}}_{X^j}$ in the condition $R=C=4$. Note that fEM is the fuzzy-EM algorithm whereas dML-max and dML-mean denote the standard maximum likelihood based on max-based and mean-based defuzzified counts.} 
	\label{tab2c}
\end{table}

\begin{table}[h!]
	\centering
	\begin{tabular}{lccccccc}
		\toprule
		\multirow{2}{*}{} &
		\multicolumn{2}{c}{fEM} &
		\multicolumn{2}{c}{dML-max} &
		\multicolumn{2}{c}{dML-mean} \\ \cmidrule(lr){2-3} \cmidrule(lr){4-5} \cmidrule(lr){6-7} 
		$R=C=6$& {\textit{bias}}& {\textit{rmse}}& {\textit{bias}}& {\textit{rmse}}& {\textit{bias}}& {\textit{rmse}}\\
		\midrule
		$\rho=0.15$ &&&&&&&\\[0.1cm]
		$I=150$ & 0.07773 & 0.32426 & 0.06858 & 0.15506 & 0.14761 & 0.17434 \\ 
		$I=250$ & 0.04557 & 0.10842 & 0.02624 & 0.09560 & 0.06524 & 0.09514 \\ 
		$I=500$ & 0.04214 & 0.06111 & 0.00563 & 0.05203 & 0.00926 & 0.04363 \\ 
		$I=1000$ & 0.01893 & 0.02878 & 0.00156 & 0.02522 & 0.00456 & 0.02381 \\[0.2cm]
		$\rho=0.50$ &&&&&&&\\[0.1cm] 
		$I=150$ & 0.13719 & 0.43117 & 0.06764 & 0.15797 & 0.14535 & 0.17206 \\ 
		$I=250$ & 0.06769 & 0.13028 & 0.02734 & 0.09601 & 0.06286 & 0.09325 \\ 
		$I=500$ & 0.01777 & 0.04274 & 0.01712 & 0.04958 & 0.02562 & 0.04733 \\ 
		$I=1000$ & 0.02963 & 0.03633 & 0.00881 & 0.02693 & 0.00994 & 0.02522 \\[0.2cm]
		$\rho=0.85$ &&&&&&&\\[0.1cm] 
		$I=150$ & 0.02021 & 0.17906 & 0.10739 & 0.15707 & 0.16366 & 0.18348 \\ 
		$I=250$ & 0.01492 & 0.06809 & 0.06039 & 0.10015 & 0.09371 & 0.11295 \\ 
		$I=500$ & 0.02625 & 0.04338 & 0.01779 & 0.04829 & 0.03246 & 0.04939 \\ 
		$I=1000$ & 0.02293 & 0.02893 & 0.00222 & 0.02407 & 0.00696 & 0.02142 \\
		\hline\bottomrule
	\end{tabular}
	\caption{Simulation study: Average bias and root mean square errors for the aggregated thresholds $\widehat{\boldsymbol{\tau}} = \mathbf 1_{R_d}^T\widehat{\boldsymbol{\tau}}_{X^j}$ in the condition $R=C=6$. Note that fEM is the fuzzy-EM algorithm whereas dML-max and dML-mean denote the standard maximum likelihood based on max-based and mean-based defuzzified counts.} 
	\label{tab2d}
\end{table}

\section{Applications}\label{sec:5}

In this section we describe the application of the proposed method to two case studies from health and natural sciences, involving the assessment of a psychotherapeutic intervention (application 1) and the evaluation of meteorological characteristics for forty Turkish cities (application 2). Note that both the applications are provided to merely illustrate the use of fuzzy LLCs model when dealing with imprecise data. 

\subsection{Application 1: Assessing the outcome of a therapy}\label{sec:5_1}

\begin{figure}[!h]
	\hspace{-1cm}
	\resizebox{18cm}{!}{\input{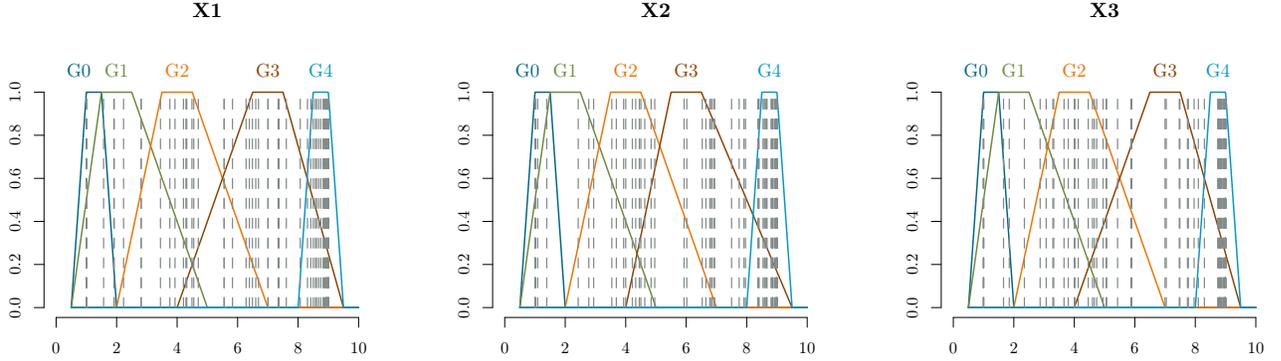}}
	\caption{Application 1: Granulation for the three fuzzy variables along with crisp observations (dashed gray lines).}
	\label{fig2}
\end{figure}

Evaluating the quality of a psychotherapy session plays a central role in evidence-based medicine. A typical approach to understand the fundamentals of the therapeutic process consists in asking experts to assess the global quality and characteristics of the therapist-patient relationship through specialized instruments such as the PQS questionnaire \cite{price1998examining}. The data thus collected generally consist either of ratings or classification of attributes made through bounded and graded scales. Because of their characteristics, these tasks often involve imprecision and vagueness that can adequately be accounted for by the fuzzy statistical modeling. In this application, we consider the assessment of a psychotherapy session by means of the PQS questionnaire. Data were originally collected by \cite{ciavolino2014fuzzy} an refers to $I=60$ evaluations of psychotherapy on a 9-point scale over $J=3$ dimensions of assessment. Given the nature of the task,
the three variables were originally considered to be fuzzy, each with three trapezoidal fuzzy categories. To account for the extremes of the classification scale, two more outer categories were added so that $R=C=5$ (see Table \ref{tab3a}). Figure \ref{fig2} shows the granulation based on five fuzzy categories ($G_0$,$\ldots$,$G_4$) for each dimensions of assessment along with the corresponding crisp observations. The aim is to compute the correlation matrix for the three fuzzy variables, with the hypothesis that the higher degree of association is related to a good therapeutic outcome. The first step requires computing the fuzzy frequency matrix $\mathbf{\widetilde N}_{5\times 5}$ for each pair of $J=3$ fuzzy variables given the crisp observed data. Next, the matrix of fuzzy counts is used to estimate the latent linear correlation matrix $\mathbf{\widehat R}_{5\times 5}$. Figure \ref{fig3} shows a graphical representation of the matrix of fuzzy counts $\mathbf{\widetilde{N}}_{5\times 5}$ for one pair of variables (i.e., $X_2$,$X_3$). It contains fuzzy numbers with various degree of fuzziness and include combinations with degenerated fuzzy counts as well (i.e., $G_0^{(2)},G_4^{(3)}$ and $G_0^{(2)},G_0^{(3)}$). Table \ref{tab3b} reports the estimates of LLC coefficients. Overall, the results showed a low level of association among the three dimensions, which in turn indicated that the psychotherapy being assessed cannot be classified as having a good outcome.

\begin{table}[h!]
	\centering
	\begin{tabular}{rcccccccccccc}
		\toprule
		\multirow{2}{*}{} &
		\multicolumn{4}{c}{$X_1$} &
		\multicolumn{4}{c}{$X_2$} &
		\multicolumn{4}{c}{$X_3$} \\
		\cmidrule(lr){2-5} \cmidrule(lr){6-9} \cmidrule(lr){10-13}
		& $x_l$ & $c_1$ & $c_2$ & $x_u$ & $x_l$ & $c_1$ & $c_2$ & $x_u$ & $x_l$ & $c_1$ & $c_2$ & $x_u$ \\ 
		\midrule
		$r=1$ & 0.50 & 1.00 & 1.50 & 2.00 & 0.50 & 1.00 & 1.50 & 2.00 & 0.50 & 1.00 & 1.50 & 2.00 \\ 
		$r=2$ & 0.50 & 1.50 & 2.50 & 5.00 & 0.50 & 1.50 & 2.50 & 5.00 & 0.50 & 1.50 & 2.50 & 5.00 \\ 
		$r=3$ & 2.00 & 3.50 & 4.50 & 7.00 & 2.00 & 3.50 & 4.50 & 7.00 & 2.00 & 3.50 & 4.50 & 7.00 \\ 
		$r=4$ & 4.00 & 6.50 & 7.50 & 9.50 & 4.00 & 5.50 & 6.50 & 9.50 & 4.00 & 6.50 & 7.50 & 9.50 \\ 
		$r=5$ & 8.00 & 8.50 & 9.00 & 9.50 & 8.00 & 8.50 & 9.00 & 9.50 & 8.00 & 8.50 & 9.00 & 9.50 \\ 
		\bottomrule
	\end{tabular}
	\caption{Application 1: Fuzzy categories for the three variables of the assessment task. Note that each category is represented by means of trapezoidal fuzzy numbers (see Eq. \ref{eq1}).}
	\label{tab3a}
\end{table}

\begin{figure}[!h]
	\centering
	\resizebox{14cm}{!}{\input{fig3.tex}}
	\caption{Application 1: Fuzzy frequency matrix for the pair $X_2$,$X_3$. Note that each cell contains a fuzzy natural number $\tilde n_{rc}$ for a specific combination of the $R\times C$ granulation space.}
	\label{fig3}
\end{figure}

\begin{table}[h!]
	\centering
	\begin{tabular}{l|ccc}
		& $X_1$ & $X_2$ & $X_3$ \\ 
		\hline
		$X_1$ & 1.00000 &  & \\ 
		$X_2$ & 0.06948 & 1.00000 & \\ 
		$X_3$ & 0.00004 & 0.21762 & 1.00000 \\ 
	\end{tabular}
	\caption{Application 1: Latent linear correlation matrix estimated via Olsson's two-stage fuzzy-EM algorithm.} 
	\label{tab3b}
\end{table}

\subsection{Application 2: Effect of climatic variables on rainfall}\label{sec:5_2}

Meteorological variables are generally used to assess the impact of climatic characteristics in many phenomena including human as well as non-human activities. Although often regarded as discrete or continuous measurements, these variables can benefit from fuzzy coding in some circumstances. Examples include cases in which these variables are imprecisely coded (e.g., when data are available in terms of intervals or linguistic categories) or when they are derived from a variety of sources (e.g., samples, historical databases, experts) that need to be integrated before being used for data analysis \cite{blasius2014visualization,chevene1994fuzzy}. In this application, we consider the analysis of $J=5$ meteorological variables (i.e., \texttt{SUN}: daily hours of sunshine; \texttt{HUM}: percentage of humidity; \texttt{PRE}: precipitations; \texttt{ALT}: altitude; \texttt{MAX}: maximum daily temperature) which were collected in forty cities of Turkey during 2004 \cite{asan2008measures}. Data were originally coded using $R=C=3$ fuzzy triangular categories ($G_0$: \textit{minimum}; $G_1$: \textit{medium}; $G_2$: \textit{maximum}) and membership grades $\boldsymbol{\epsilon}^{(j)}_1,\boldsymbol{\epsilon}^{(j)}_2,\boldsymbol{\epsilon}^{(j)}_3$ $j=1,\ldots,5$ constitute the input data for the subsequent analysis. The aim is to explore the effects of climatic variables on rainfall (\texttt{PRE}) by means of a path analysis model. Likewise for the first application, the first step consisted in computing the fuzzy frequency matrix $\mathbf{\widehat N}_{3\times 3}$ for each pair of the five climatic variables given the observed membership degrees. Then, the LLCs matrix was estimated using the fuzzy-EM algorithm. Figure \ref{fig4} shows an example of fuzzy counts for the pair of variables \texttt{PRE}-\texttt{HUM} whereas Table \ref{tab4a} reports the estimated correlations for the variables involved in the study. As expected, the results showed a certain level of association among the five climatic variables.  

\begin{figure}[!h]
	\centering
	\resizebox{11cm}{!}{\input{fig4.tex}}
	\caption{Application 2: Fuzzy frequency matrix for the pair \texttt{PRE}, \texttt{HUM}. Note that each cell contains a fuzzy natural number $\tilde n_{rc}$ for a specific combination of the $R\times C$ granulation space.}
	\label{fig4}
\end{figure}

\begin{table}[h!]
	\centering
	\begin{tabular}{l|ccccc}
		& \texttt{SUN} & \texttt{HUM} & \texttt{PRE} & \texttt{ALT} & \texttt{MAX} \\ 
		\hline
		\texttt{SUN} & 1.00000 & & & &  \\ 
		\texttt{HUM} & -0.73125 & 1.00000 & & & \\ 
		\texttt{PRE} & -0.58327 & 0.31726 & 1.00000 & & \\ 
		\texttt{ALT} & -0.21412 & -0.47867 & -0.92587 & 1.00000 & \\ 
		\texttt{MAX} & 0.43941 & -0.23214 & 0.17976 & -0.58675 & 1.00000 \\ 
	\end{tabular}
	\caption{Application 2: Latent linear correlation matrix estimated via Olsson's two-stage fuzzy-EM algorithm.} 
	\label{tab4a}
\end{table}

\begin{figure}[!h]
	\centering
	\resizebox{10cm}{!}{
\begin{tikzpicture}[x=1pt,y=1pt]
\definecolor{fillColor}{RGB}{255,255,255}
\path[use as bounding box,fill=fillColor,fill opacity=0.00] (0,0) rectangle (614.29,289.08);
\begin{scope}
\path[clip] (  0.00,  0.00) rectangle (614.29,289.08);
\definecolor{drawColor}{gray}{0.30}

\path[draw=drawColor,line width= 1.6pt,line join=round,line cap=round] (391.19,224.84) --
	(484.48,153.64);
\definecolor{fillColor}{gray}{0.30}

\path[fill=fillColor] (482.47,151.01) --
	(490.82,148.79) --
	(486.48,156.27) --
	cycle;

\path[draw=drawColor,line width= 1.6pt,line join=round,line cap=round] (299.63, 64.24) --
	(483.22,123.44);

\path[fill=fillColor] (484.24,120.29) --
	(490.82,125.89) --
	(482.21,126.59) --
	cycle;

\path[draw=drawColor,line width= 1.6pt,line join=round,line cap=round] (121.23,205.36) --
	(362.14,222.74);

\path[fill=fillColor] (362.38,219.44) --
	(370.11,223.32) --
	(361.90,226.04) --
	cycle;

\path[draw=drawColor,line width= 1.6pt,line join=round,line cap=round] (102.38,100.93) --
	(116.01,176.42);

\path[fill=fillColor] (119.26,175.83) --
	(117.43,184.28) --
	(112.75,177.00) --
	cycle;

\path[draw=drawColor,line width= 1.6pt,dash pattern=on 4pt off 4pt ,line join=round,line cap=round] (270.69, 69.62) --
	(131.32, 95.55);

\path[fill=fillColor] (131.93, 98.80) --
	(123.47, 97.01) --
	(130.72, 92.29) --
	cycle;

\path[fill=fillColor] (270.09, 66.37) --
	(278.54, 68.16) --
	(271.30, 72.88) --
	cycle;
\definecolor{fillColor}{RGB}{238,118,0}

\path[fill=fillColor] (490.82,111.60) --
	(533.00,111.60) --
	(533.00,153.78) --
	(490.82,153.78) --
	cycle;

\path[] (490.82,111.60) --
	(533.00,111.60) --
	(533.00,153.78) --
	(490.82,153.78) --
	(490.82,111.60);
\definecolor{drawColor}{RGB}{0,0,0}

\path[draw=drawColor,line width= 0.4pt,line join=round,line cap=round] (490.82,111.60) rectangle (533.00,153.78);
\definecolor{fillColor}{RGB}{162,205,90}

\path[fill=fillColor] (370.11,203.75) --
	(412.28,203.75) --
	(412.28,245.93) --
	(370.11,245.93) --
	cycle;

\path[] (370.11,203.75) --
	(412.28,203.75) --
	(412.28,245.93) --
	(370.11,245.93) --
	(370.11,203.75);

\path[draw=drawColor,line width= 0.4pt,line join=round,line cap=round] (370.11,203.75) rectangle (412.28,245.93);

\path[fill=fillColor] (100.15,184.28) --
	(142.32,184.28) --
	(142.32,226.45) --
	(100.15,226.45) --
	cycle;

\path[] (100.15,184.28) --
	(142.32,184.28) --
	(142.32,226.45) --
	(100.15,226.45) --
	(100.15,184.28);

\path[draw=drawColor,line width= 0.4pt,line join=round,line cap=round] (100.15,184.28) rectangle (142.32,226.45);

\path[fill=fillColor] (278.54, 43.15) --
	(320.72, 43.15) --
	(320.72, 85.33) --
	(278.54, 85.33) --
	cycle;

\path[] (278.54, 43.15) --
	(320.72, 43.15) --
	(320.72, 85.33) --
	(278.54, 85.33) --
	(278.54, 43.15);

\path[draw=drawColor,line width= 0.4pt,line join=round,line cap=round] (278.54, 43.15) rectangle (320.72, 85.33);

\path[fill=fillColor] ( 81.29, 79.84) --
	(123.47, 79.84) --
	(123.47,122.02) --
	( 81.29,122.02) --
	cycle;

\path[] ( 81.29, 79.84) --
	(123.47, 79.84) --
	(123.47,122.02) --
	( 81.29,122.02) --
	( 81.29, 79.84);

\path[draw=drawColor,line width= 0.4pt,line join=round,line cap=round] ( 81.29, 79.84) rectangle (123.47,122.02);

\node[text=drawColor,anchor=base,inner sep=0pt, outer sep=0pt, scale=  1.38] at (511.91,127.93) {Pre};

\node[text=drawColor,anchor=base,inner sep=0pt, outer sep=0pt, scale=  1.38] at (391.19,220.08) {Hum};

\node[text=drawColor,anchor=base,inner sep=0pt, outer sep=0pt, scale=  1.38] at (121.23,200.60) {Max};

\node[text=drawColor,anchor=base,inner sep=0pt, outer sep=0pt, scale=  1.38] at (299.63, 59.48) {Sun};

\node[text=drawColor,anchor=base,inner sep=0pt, outer sep=0pt, scale=  1.38] at (102.38, 96.17) {Alt};
\end{scope}
\end{tikzpicture}}
	\caption{Application 2: Path model for the effect of the climatic variables on the response variable \texttt{PRE}. Note that straight lines represent direct effects whereas dotted lines indicate correlations.}
	\label{fig5}
\end{figure}
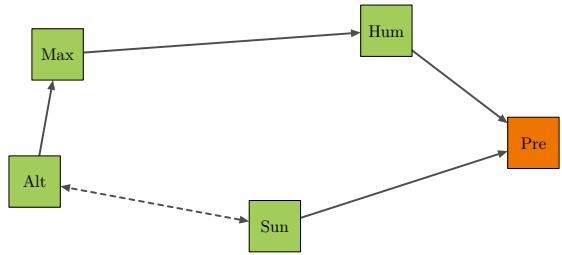

Once the LLCs matrix has been estimated, we proceeded by modeling the effects of the climatic variables on \texttt{PRE} via path analysis (see Figure \ref{fig5}). In particular, we expected that a higher humidity (\texttt{HUM}) increased rainfall (\texttt{PRE}) and that sunshine duration (\texttt{SUN}) decreased the levels of precipitation (\texttt{PRE}). Similarly, we also expected an indirect effect of altitude (\texttt{ALT}) on humidity (\texttt{HUM}) through temperatures (\texttt{TEMP}). The path model has been estimated on the LLCs matrix via maximum likelihood as implemented in the \texttt{R} library \texttt{lavaan} \cite{rosseel2012lavaan}. Overall, the estimated model showed a moderate fit ($R^2=0.20$). The results highlighted that \texttt{PRE} increased as a function of \texttt{HUM} ($\hat{\beta}=0.1844$, $\hat{\sigma}^2_\beta=0.1386$) and descreased as sunshine duration increased ($\hat{\beta}=-0.3406$, $\hat{\sigma}^2_\beta=0.1386$). Humidity was negatively related to temperature ($\hat{\beta}=-0.2161$, $\hat{\sigma}^2_\beta=0.1544$), which was in turn negatively associated to altitude ($\hat{\beta}=-0.5577$, $\hat{\sigma}^2_\beta=0.1312$) as expected.

\begin{table}[h!]
	\centering
	\begin{tabular}{lccc}
		\toprule
		path & $\hat\beta$ & $\hat\sigma^2_{\beta}$ & $\hat\sigma^2_\epsilon$ \\ 
		\midrule
		\texttt{HUM}-\texttt{PRE} & 0.1844 & 0.1386 & 0.7488 \\ 
		\texttt{SUN}-\texttt{PRE} & -0.3406 & 0.1386 & 0.9295\\ 
		\texttt{MAX}-\texttt{HUM} & -0.2161 & 0.1544 & 0.6717\\ 
		\texttt{ALT}-\texttt{MAX} & -0.5577 & 0.1312 & 0.975\\ 
		\bottomrule
	\end{tabular}
	\caption{Application 2: Estimated coefficients $\boldsymbol{\hat\beta}$ and residual variances $\boldsymbol{\hat\sigma}_\epsilon^2$ for the path model depicted in Figure \ref{fig5} along with the standard errors $\boldsymbol{\hat\sigma}_\beta^2$ of the estimates. } 
	\label{tab4b}
\end{table}

\section{Conclusions}\label{sec:6}

In this article we described a novel approach to estimate latent linear correlations (LLCs) when data are in the form of fuzzy frequency tables. In particular, we represented fuzzy counts in terms of generalized natural numbers first, and then we generalized the sample space of the standard LLCs model to cope with fuzzy counts while retaining its parameter space as non-fuzzy. The resulting model encapsulated both random and non-random/imprecision components in a unified statistical representation. Since the inferential interest is on estimating the latent correlation matrix of the observed variables, parameter estimation was performed via fuzzy maximum likelihood using the Expectation-Maximization algorithm. A simulation study and two real applications were developed to highlight the characteristics of the fuzzy LLCs model. Overall, the simulation results revealed that the fuzzy LLCs model showed more accurate results in estimating the true correlation matrix as opposed to standard methods which can be applied on defuzzified data. The applications showed how the proposed method can be of particular value in situations involving fuzzy classification and fuzzy coding as well.

A particular advantage of the fuzzy LLCs model is its simplicity and ability to deal with situations involving imprecise classification problems. Moreover, the proposed method works with both fuzzy observations/crisp categories and crisp observations/fuzzy categories and, as such, it includes the standard crisp observations/crisp categories as a special case. Again, the fuzzy LLCs model does not require the extension of its parametric representation to account for fuzzy frequency data and, consequently, parameter estimation and inference can be performed using the asymptotic properties of maximum likelihood theory. This is quite convenient and obviates the need of generalizing LLCs-based statistical modeling - such as structural equation models and factor analysis - to the fuzzy case. A limitation of the proposed approach is that it is based on the simplest, but still used, assumption of Gaussianity for LLCs. Although it has been proved that the assumption holds in several empirical contexts, there may be the need of LLCs based on more general probabilistic models (e.g., Skew-Gaussian, Elliptical, $t$, Copula-based). As a result, the problems already identified by other researchers, for instance bias in estimating the asymptotic covariance matrix of the LLCs matrix \cite{monroe2018contributions,foldnes2020pernicious}, still persist in the fuzzy case. 

There are a
number of further extensions to this project that can be undertaken in future research studies. For instance, the use of more general probabilistic model would extend the proposed method to handle with situations involving violations of Gaussianity assumption. Another aspect which might be interesting to investigate is the case where data need to be represented using more general fuzzy numbers (e.g., beta, exponential, gaussian), which would allow the proposed method to cope with cases requiring more flexible models to represent non-random imprecision. Finally, studying the properties of fuzzy LLCs-based statistical models like structural equation modeling or factor analysis would also constitute a research topic to be considered in a further study.

\vspace{1.5cm}
\noindent\textbf{Acknowledgments}. The author wishes to acknowledge and thank Dr. Andrea Spirito for his valuable comments on various issues concerning this research study.

\clearpage
\renewcommand{\theequation}{A.\arabic{equation}} 
\setcounter{equation}{0}  
\begin{appendix}

\section{Appendix}\label{apx1}

To compute the nonlinear expectation $\Expp{\boldsymbol{\theta}'}{\ln N_{rc}^{jk}! \Big| \tilde{n}^{jk}_{rc}}$, we first approximate the factorial term via Stirling's formula:
\begin{align*}
	\Expp{\boldsymbol{\theta}'}{\ln N_{rc}^{jk}! \Big| \tilde{n}^{jk}_{rc}} &= \Expp{\boldsymbol{\theta}'}{ N_{rc}^{jk} \ln N_{rc}^{jk} - N_{rc}^{jk} + \frac{1}{2}\ln2\pi \Big| \tilde{n}^{jk}_{rc}} \nonumber\\
	& = \Expp{\boldsymbol{\theta}'}{ N_{rc}^{jk} \ln N_{rc}^{jk} \Big| \tilde{n}^{jk}_{rc}} - \Expp{\boldsymbol{\theta}'}{ N_{rc}^{jk} \Big| \tilde{n}^{jk}_{rc}} \nonumber\\
	& = \Expp{\boldsymbol{\theta}'}{ g\left( N_{rc}^{jk} \right) \Big| \tilde{n}^{jk}_{rc}} - \Expp{\boldsymbol{\theta}'}{ N_{rc}^{jk} \Big| \tilde{n}^{jk}_{rc}} \nonumber
\end{align*}
with $g(x):= x\ln x$. Next, since the non linear transformation $g(.)$ is smooth and twice-differentiable on $(0,\infty)$ with $g''(x) = 1/x$, a second-order Taylor expansion around the first conditional moment $\Expp{\boldsymbol{\theta}'}{N_{rc}^{jk}\Big| {\tilde{n}_{rc}}}$ can be developed to get the closed-form expression of the expectation term:
\begin{equation}\label{eqA1}
	 \Expp{\boldsymbol{\theta}'}{g(N_{rc}^{jk})\Big| {\tilde{n}_{rc}}^{jk}}\approxx g\left(\Expp{\boldsymbol{\theta}'}{N_{rc}^{jk}\Big| {\tilde{n}_{rc}}^{jk}}\right) + \frac{\Varr{\boldsymbol{\theta}'}{N_{rc}^{jk}\Big| {\tilde{n}_{rc}}^{jk}}}{2\Expp{\boldsymbol{\theta}'}{N_{rc}^{jk}\Big| {\tilde{n}_{rc}}^{jk}}} 
\end{equation}
with the conditional variance being defined by
\begin{equation*}
	\Varr{\boldsymbol{\theta}'}{N_{rc}^{jk}\Big| {\tilde{n}_{rc}}^{jk}} = \sum_{n\in\mathbb N_0} \left(n-\Expp{\boldsymbol{\theta}'}{N_{rc}^{jk}\Big| {\tilde{n}_{rc}}^{jk}}\right)^2 p_{N_{rc}^{jk}|\tilde{n}_{rc}^{jk}}(n;\pi_{rc}^{jk}(\boldsymbol{\theta}'))	
\end{equation*} 
where $\Expp{\boldsymbol{\theta}'}{N_{rc}^{jk}\Big| {\tilde{n}_{rc}}^{jk}}$ is as in Eq. \eqref{eq9a}.

\section{Appendix}\label{apx2}

To establish monotonicity for a sequence of log-likelihood evaluations $\{\ln \mathcall L(\boldsymbol{\theta}^{(q)};\mathbf{\tilde N})\}_{q\in \mathbb{N}}$ of the fuzzy Expectation Maximization algorithm we will follow the general results of \cite{mclachlan2007algorithm}, Sect. 3.2. A similar proof is also given by \cite{su2015parameter} for the case of rectangular fuzzy numbers (i.e., interval-valued data). In what follows, we will omit the indices $j,k$ for the sake of simplicity. Given $\boldsymbol{\theta}' = \boldsymbol{\theta}^{(q-1)}$ and by rearranging Eq. \eqref{eq8}, we get by standard calculus:
\begin{align}
	\ln\mathcal L(\boldsymbol{\theta};\widetilde{\mathbf{N}}) &= \ln \mathcall L(\boldsymbol{\theta};\mathbf{N}) -\sum_{r=1}^R \sum_{c=1}^C \ln p_{N_{rc}|\tilde{n}_{rc}}(n;\pi_{rc}(\boldsymbol{\theta})) \nonumber\\
	& =\Expp{\boldsymbol{\theta}'}{\sum_{r=1}^R\sum_{c=1}^C \ln \mathcall L(\boldsymbol{\theta};{N}_{rc}) \Big|\tilde n_{rc}} - \Expp{\boldsymbol{\theta}'}{\sum_{r=1}^R \sum_{c=1}^C \ln p_{N_{rc}|\tilde n_{rc}}(n;\pi_{rc}(\boldsymbol{\theta}))\Big|\tilde{n}_{rc}}\nonumber\\
	&= \mathcall Q(\boldsymbol{\theta};\boldsymbol{\theta}') - \mathcall S(\boldsymbol{\theta};\boldsymbol{\theta}')\label{eqA2}
\end{align}
Then, an increasing of the observed log-likelihood can be written in terms of the result \eqref{eqA2} as follows:
\begin{align*}
	\ln \mathcall L(\boldsymbol{\theta}^{(q)};\mathbf{\widetilde N}) - \ln \mathcall L(\boldsymbol{\theta}';\mathbf{\widetilde N}) ~\geq&~ \left(\mathcall Q(\boldsymbol{\theta}^{(q)};\boldsymbol{\theta}'))-\mathcall Q(\boldsymbol{\theta}';\boldsymbol{\theta}')\right) - \left( \mathcall S(\boldsymbol{\theta}^{(q)};\boldsymbol{\theta}') - \mathcall S(\boldsymbol{\theta}';\boldsymbol{\theta}') \right)
\end{align*}
Note that because $\boldsymbol{\theta}^{(q)}$ is chosen so that $\mathcall Q(\boldsymbol{\theta}^{(q)};\boldsymbol{\theta}'))-\mathcall Q(\boldsymbol{\theta}';\boldsymbol{\theta}') \geq 0$ \cite{mclachlan2007algorithm}, the condition $\mathcall S(\boldsymbol{\theta};\boldsymbol{\theta}') - \mathcall S(\boldsymbol{\theta}';\boldsymbol{\theta}') \leq 0$ must hold for each $\boldsymbol{\theta}$. To do so, we proceed as follows:
\begin{align*}
	\mathcall S(\boldsymbol{\theta};\boldsymbol{\theta}') - \mathcall S(\boldsymbol{\theta}';\boldsymbol{\theta}') & = \sum_{r=1}^R \sum_{c=1}^C \Expp{\boldsymbol{\theta}'}{\ln p_{N_{rc}|\tilde n_{rc}}(n;\pi_{rc}(\boldsymbol{\theta}))\Big|\tilde n_{rc}} - \Expp{\boldsymbol{\theta}'}{\ln p_{N_{rc}|\tilde n_{rc}}(n;\pi_{rc}(\boldsymbol{\theta}'))\Big|\tilde n_{rc}}\\
	& = \sum_{r=1}^R \sum_{c=1}^C  \Expp{\boldsymbol{\theta}'}{\ln \left( \frac{p_{N_{rc}|\tilde n_{rc}}(n;\pi_{rc}(\boldsymbol{\theta}))}{p_{N_{rc}|\tilde n_{rc}}(n;\pi_{rc}(\boldsymbol{\theta}'))} \right) \Bigg|\tilde n_{rc}}\\
	& \leq \sum_{r=1}^R \sum_{c=1}^C  \ln \Expp{\boldsymbol{\theta}'}{\frac{p_{N_{rc}|\tilde n_{rc}}(n;\pi_{rc}(\boldsymbol{\theta}))}{p_{N_{rc}|\tilde n_{rc}}(n;\pi_{rc}(\boldsymbol{\theta}'))} \Bigg|\tilde n_{rc}}\quad\text{\footnotesize using Jensen's inequality}\\
	& \leq \sum_{r=1}^R \sum_{c=1}^C ~ \ln \sum_{n\in\mathbb N_0} \frac{p_{N_{rc}|\tilde n_{rc}}(n;\pi_{rc}(\boldsymbol{\theta}))}{p_{N_{rc}|\tilde n_{rc}}(n;\pi_{rc}(\boldsymbol{\theta}'))} p_{N_{rc}|\tilde n_{rc}}(n;\pi_{rc}(\boldsymbol{\theta}')) ~n \\
	& \leq \sum_{r=1}^R \sum_{c=1}^C ~ \ln \sum_{n\in\mathbb N_0} {p_{N_{rc}|\tilde n_{rc}}(n;\pi_{rc}(\boldsymbol{\theta}))} ~n\\
	& \leq \sum_{r=1}^R \sum_{c=1}^C ~ \ln 1 = 0 \qed
\end{align*}
Hence, an increasing of $\ln \mathcall L(\boldsymbol{\theta}^{(q)};\mathbf{\widetilde N}) - \ln \mathcall L(\boldsymbol{\theta}';\mathbf{\widetilde N}) \geq 0$ is guaranteed as soon as $\mathcall Q(\boldsymbol{\theta}^{(q)};\boldsymbol{\theta}'))-\mathcall Q(\boldsymbol{\theta}';\boldsymbol{\theta}') \geq 0$.	

\end{appendix}

\clearpage
\bibliographystyle{unsrt} 
\bibliography{biblio}

\clearpage
%
%
%
%


\begin{center}
	\Large \underline{Supplementary Materials}\\\vspace{1cm} \normalsize
\end{center}		


\renewcommand{\thetable}{S.\arabic{table}}\setcounter{table}{0}  
\renewcommand{\thefigure}{S.\arabic{figure}}\setcounter{figure}{0}  

\begin{table}[h!]
	\centering
	\begin{tabular}{lcccccccc}
		\toprule
		\multirow{2}{*}{} &
		\multicolumn{3}{c}{$R=C=4$} &
		\multicolumn{5}{c}{$R=C=6$} \\
		\cmidrule(lr){2-4} \cmidrule(lr){5-9}
		& {$\hat\tau_1$}& {$\hat\tau_2$}& {$\hat\tau_3$} & {$\hat\tau_1$}& {$\hat\tau_2$}& {$\hat\tau_3$} & {$\hat\tau_4$}& {$\hat\tau_5$}\\
		\midrule
		$\rho=0.15$ &&&&&&&&\\[0.1cm]
		$I=150$ & -2.15645 & 0.01727 & 2.15876 & -2.18341 & -1.00894 & -0.00060 & 1.00818 & 2.18752 \\ 
		$I=250$ & -2.10902 & 0.02070 & 2.10671 & -2.06642 & -1.03889 & 0.02140 & 1.03830 & 2.06281 \\ 
		$I=500$ & -2.03274 & 0.00006 & 2.03286 & -2.07933 & -1.02383 & 0.00543 & 1.02429 & 2.07785 \\ 
		$I=1000$ & -2.04243 & 0.00261 & 2.04244 & -2.03466 & -1.01247 & -0.00002 & 1.01250 & 2.03503 \\[0.2cm]
		$\rho=0.50$ &&&&&&&&\\[0.1cm] 
		$I=150$ & -1.99019 & -0.00017 & 1.99206 & -2.32291 & -1.01693 & 0.00006 & 1.01759 & 2.32844 \\ 
		$I=250$ & -2.06659 & -0.00030 & 2.06650 & -2.09860 & -1.05156 & 0.03270 & 1.05188 & 2.10372 \\ 
		$I=500$ & -2.06413 & 0.00510 & 2.06411 & -2.03232 & -1.01273 & -0.00017 & 1.01209 & 2.03153 \\ 
		$I=1000$ & -2.02622 & 0.00005 & 2.02687 & -2.05360 & -1.01820 & 0.00520 & 1.01829 & 2.05286 \\[0.2cm]
		$\rho=0.85$ &&&&&&&&\\[0.1cm] 
		$I=150$ & -2.09765 & 0.00008 & 2.09203 & -2.02365 & -1.02091 & 0.01715 & 1.01890 & 2.02043 \\ 
		$I=250$ & -2.05866 & -0.00023 & 2.05640 & -2.02009 & -1.01721 & 0.00020 & 1.01737 & 2.01975 \\ 
		$I=500$ & -2.02893 & -0.00021 & 2.02840 & -2.04960 & -1.01681 & -0.00013 & 1.01626 & 2.04844 \\ 
		$I=1000$ & -2.04424 & 0.00249 & 2.04435 & -2.04356 & -1.01229 & 0.00263 & 1.01250 & 2.04368 \\ 		
		\hline\bottomrule
	\end{tabular}
	\caption{Simulation study: Average mean values of $\widehat{\boldsymbol{\tau}}_{X^j}$ for fEM algorithm. Note that, by design, $\boldsymbol{\tau}_{X^j}=\boldsymbol{\tau}_{X^k}$.} 
	\label{tabS1}
\end{table}

\begin{table}[h!]
	\centering
	\begin{tabular}{lcccccccc}
		\toprule
		\multirow{2}{*}{} &
		\multicolumn{3}{c}{$R=C=4$} &
		\multicolumn{5}{c}{$R=C=6$} \\
		\cmidrule(lr){2-4} \cmidrule(lr){5-9}
		& {$\hat{\sigma}^2_{\hat\tau_1}$}& {$\hat{\sigma}^2_{\hat\tau_2}$}& {$\hat{\sigma}^2_{\hat\tau_3}$} & {$\hat{\sigma}^2_{\hat\tau_1}$}& {$\hat{\sigma}^2_{\hat\tau_2}$}& {$\hat{\sigma}^2_{\hat\tau_3}$} & {$\hat{\sigma}^2_{\hat\tau_4}$}& {$\hat{\sigma}^2_{\hat\tau_5}$}\\
		\midrule
		$\rho=0.15$ &&&&&&&&\\[0.1cm]
		$I=150$ & 0.15805 & 0.00080 & 0.17714 & 0.47248 & 0.00541 & 0.00202 & 0.00532 & 0.47980 \\ 
		$I=250$ & 0.01679 & 0.00027 & 0.01594 & 0.03553 & 0.00195 & 0.00069 & 0.00190 & 0.03272 \\ 
		$I=500$ & 0.00328 & 0.00007 & 0.00323 & 0.00650 & 0.00048 & 0.00018 & 0.00049 & 0.00663 \\ 
		$I=1000$ & 0.00098 & 0.00002 & 0.00097 & 0.00145 & 0.00013 & 0.00005 & 0.00013 & 0.00145 \\[0.2cm]
		$\rho=0.50$ &&&&&&&&\\[0.1cm] 
		$I=150$ & 0.02713 & 0.00083 & 0.02812 & 0.88482 & 0.00555 & 0.00217 & 0.00576 & 0.81502 \\ 
		$I=250$ & 0.01421 & 0.00029 & 0.01401 & 0.03938 & 0.00198 & 0.00075 & 0.00206 & 0.05182 \\ 
		$I=500$ & 0.00364 & 0.00007 & 0.00375 & 0.00473 & 0.00044 & 0.00017 & 0.00045 & 0.00476 \\ 
		$I=1000$ & 0.00090 & 0.00002 & 0.00085 & 0.00147 & 0.00012 & 0.00004 & 0.00012 & 0.00140 \\[0.2cm]
		$\rho=0.85$ &&&&&&&&\\[0.1cm] 
		$I=150$ & 0.22442 & 0.00081 & 0.20248 & 0.12743 & 0.00409 & 0.00159 & 0.00417 & 0.12891 \\ 
		$I=250$ & 0.01342 & 0.00028 & 0.01247 & 0.01339 & 0.00142 & 0.00055 & 0.00140 & 0.01329 \\ 
		$I=500$ & 0.00290 & 0.00007 & 0.00290 & 0.00365 & 0.00034 & 0.00013 & 0.00034 & 0.00370 \\ 
		$I=1000$ & 0.00082 & 0.00002 & 0.00079 & 0.00096 & 0.00008 & 0.00003 & 0.00009 & 0.00099 \\ 		
		\hline\bottomrule
	\end{tabular}
	\caption{Simulation study: Average variance values of $\widehat{\boldsymbol{\tau}}_{X^j}$ for fEM algorithm. Note that, by design, $\boldsymbol{\tau}_{X^j}=\boldsymbol{\tau}_{X^k}$.} 
	\label{tabS2}
\end{table}

\begin{figure}[!h]
	\centering
	\resizebox{15cm}{!}{\input{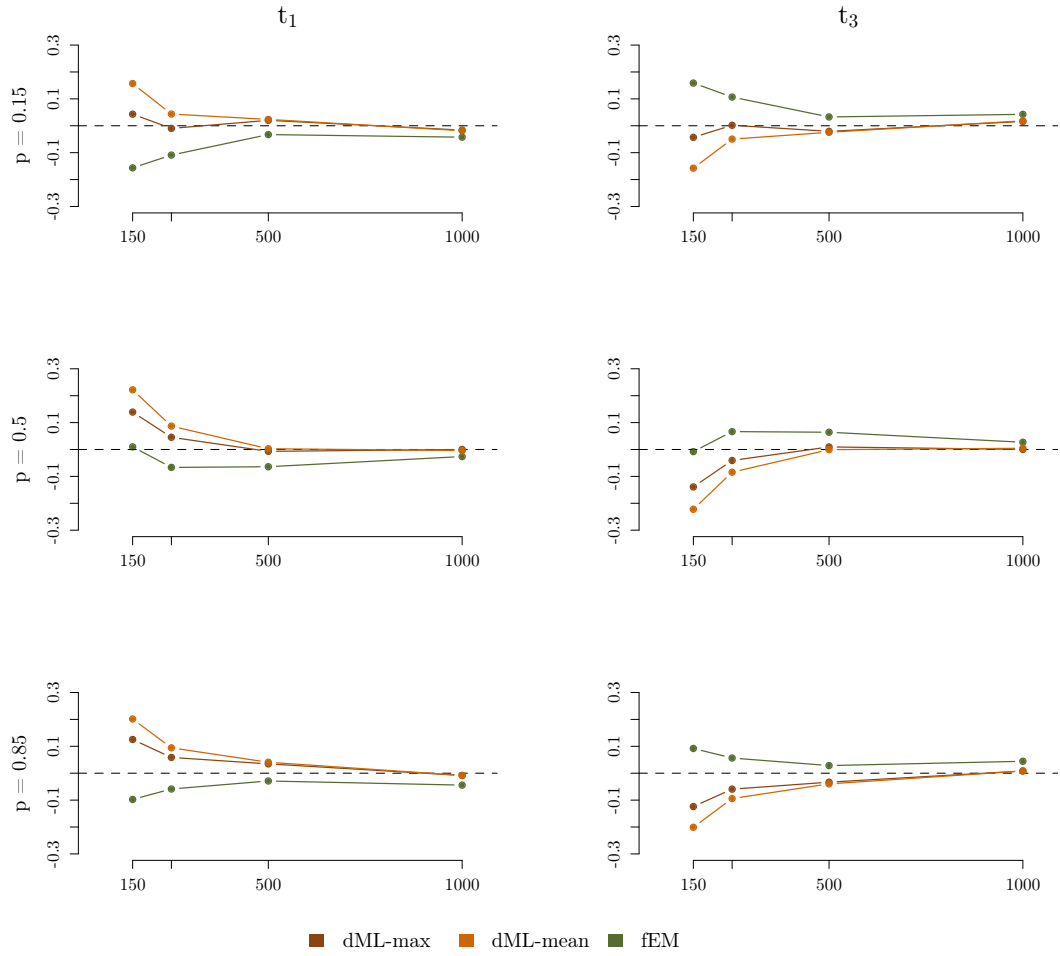}}
	\caption{Simulation study: Average bias for leftmost and rightmost values of $\boldsymbol{\tau}_{X^j}$ in the case $R=C=4$. Note that, by design, $\boldsymbol{\tau}_{X^j}=\boldsymbol{\tau}_{X^k}$.}
	\label{figS1}
\end{figure}

\begin{figure}[!h]
	\hspace{-2.5cm}
	\resizebox{20cm}{!}{\input{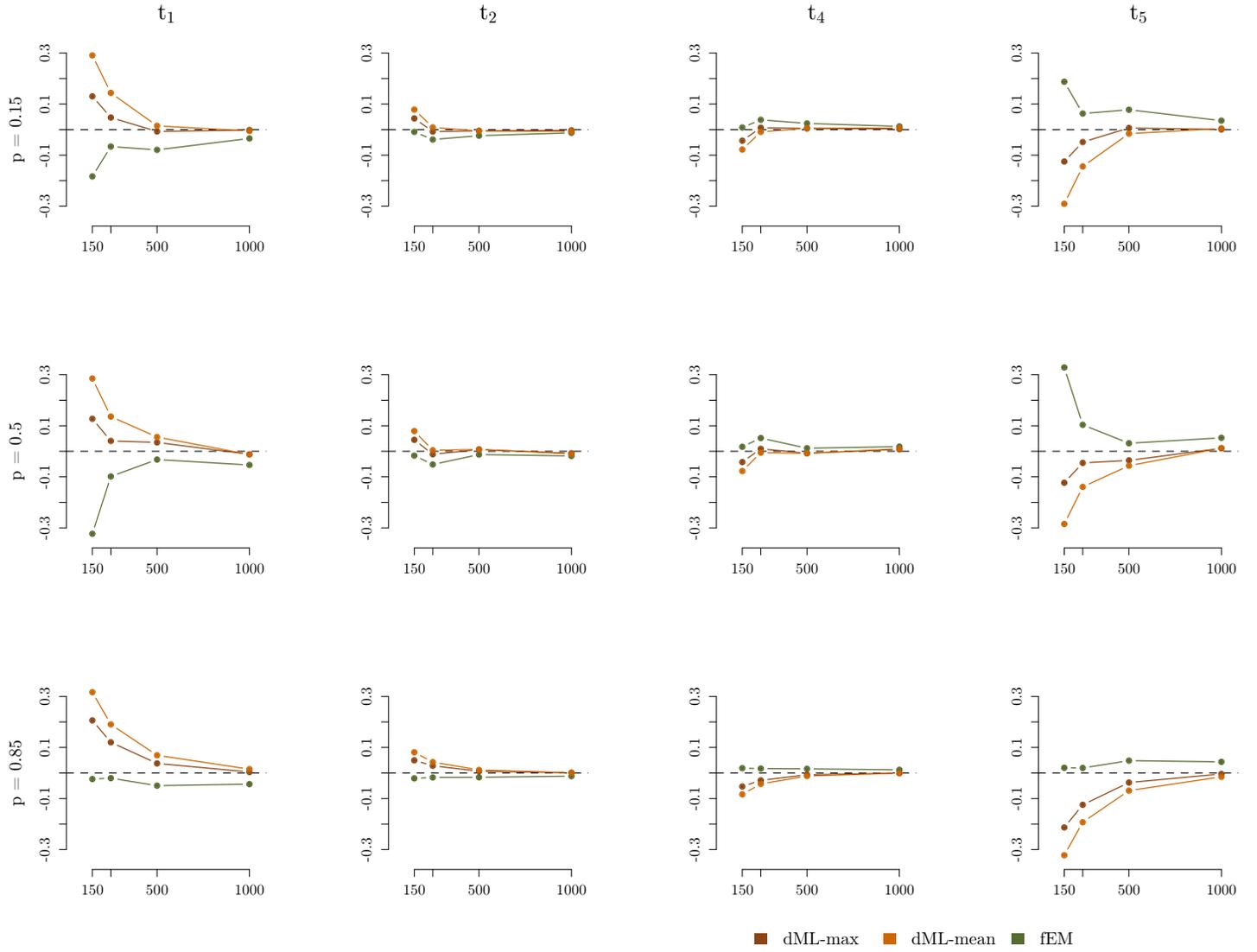}}
	\caption{Simulation study: Average bias for or leftmost and rightmost values of $\boldsymbol{\tau}_{X^j}$ in the case $R=C=6$. Note that, by design, $\boldsymbol{\tau}_{X^j}=\boldsymbol{\tau}_{X^k}$.}
	\label{figS2}
\end{figure}


\end{document}